\documentclass[12pt]{article}

\usepackage{amssymb, amsmath, amsthm, amsfonts, graphicx, color, bm, float} 
\usepackage[margin=1in]{geometry}
\usepackage{hyperref}
\usepackage{chicago}
\usepackage{setspace} 
\usepackage{graphicx, subfig, epstopdf, enumitem}
\usepackage{caption}

\usepackage{algorithm, algpseudocode}

\newfloat{algorithm}{htbp}{loa}
\floatname{algorithm}{Algorithm}

\title{Overview of Approximate Bayesian Computation}

\author{S. A. Sisson\footnote{School of Mathematics and Statistics, University of New South Wales, Sydney.}\:\:  Y. Fan$^{*}$\:\: and M. A. Beaumont\footnote{School of Mathematics, University of Bristol.}}

\begin{document}

\maketitle

\section{Introduction}

In Bayesian inference, complete knowledge about a vector of model parameters, $\theta\in\Theta$, obtained by fitting a model $\mathcal{M}$, is contained in the posterior distribution.
Here, prior beliefs about the model parameters as expressed through the prior distribution, $\pi(\theta)$, are updated by observing data $y_{obs}\in\mathcal{Y}$ through the likelihood function $\pi(y_{obs}|\theta)$ of the model. Using Bayes' Theorem, the resulting posterior distribution
\[
	 \pi(\theta|y_{obs})=\frac{p(y_{obs}|\theta)\pi(\theta)}{\int_\Theta p(y_{obs}|\theta)\pi(\theta)d\theta},
\]
contains all necessary information required for analysis of the model, including model checking and validation, predictive inference and decision making. Typically, the complexity of the model and/or prior means that the posterior distribution, $\pi(\theta|y_{obs})$, is not available in closed form, and so numerical methods are needed to proceed with the inference. A common approach makes use of Monte Carlo integration to enumerate the necessary integrals. This relies on the ability to draw samples $\theta^{(1)},\theta^{(2)},\ldots,\theta^{(N)}\sim\pi(\theta|y_{obs})$ from the posterior distribution so that a finite sample approximation to the posterior is given by the empirical measure
\[
	\pi(\theta|y_{obs})\approx \frac{1}{N}\sum_{i=1}^N\delta_{\theta^{(i)}}(\theta),
\]
where $\delta_Z(z)$ denotes the Dirac measure, defined as $\delta_Z(z)=1$ if $z\in Z$ and $\delta_Z(z)=0$ otherwise.
As the size of the sample from the posterior gets large, then the finite sample approximation better approximates the true posterior so that $\lim_{N\rightarrow\infty} \frac{1}{N}\sum_{i=1}^N\delta_{\theta^{(i)}}(\theta) \rightarrow \pi(\theta|y_{obs})$, by the law of large numbers.
As a result, the expectation of a function $a(\theta)$ under $\pi(\theta|y_{obs})$ can be estimated as
\begin{eqnarray*}
	\mathbb{E}_\pi[a(\theta)] &=&\int_\Theta a(\theta)\pi(\theta|y_{obs})d\theta\\
	& \approx & \int_\Theta a(\theta)\frac{1}{N}\sum_{i=1}^N\delta_{\theta^{(i)}}(\theta)d\theta
	= \frac{1}{N}\sum_{i=1}^Na(\theta^{(i)}).
\end{eqnarray*}

There are a number of popular algorithms available for generating samples from posterior distributions, such as importance sampling, Markov chain Monte Carlo (MCMC) and sequential Monte Carlo (SMC) \shortcite{brooksgjm11,chen+si00,doucet+dg01,delmoral+dj06}.

Inherent in such Monte Carlo algorithms is the need to numerically evaluate the posterior distribution, $\pi(\theta|y_{obs})$, up to a normalisation constant, commonly many thousands or millions of times. For example, in the Metropolis-Hastings algorithm, an MCMC algorithm, this arises through computing the probability that the Markov chain accepts the proposed move from a current point $\theta$ to a proposed point $\theta'\sim q(\theta,\theta')$ where $q$ is some proposal density, given by  $\alpha(\theta,\theta')=\min\left\{1,\frac{\pi(\theta'|y_{obs})q(\theta',\theta)}{\pi(\theta|y_{obs})q(\theta,\theta')}\right\}$. Similarly in SMC algorithms, the incremental particle weight is given by $w_t(\theta_t)=\frac{\pi_t(\theta_t|y_{obs})L_{t-1}(\theta_t,\theta_{t-1})}{\pi_{t-1}(\theta_{t-1}|y_{obs})M_t(\theta_{t-1},\theta_t)}$, where $M_t$ and $L_{t-1}$ are transition kernels, and $\pi_t$ denotes a function strongly related to the posterior distribution, such as $\pi_t(\theta_t|y_{obs})=[\pi(\theta_t|y_{obs})]^{t/T}\pi(\theta_t)^{1-t/T}$. Evaluating acceptance probabilities or particle weights clearly requires evaluation of the likelihood function.

However, for an increasing range of scientific problems -- see Section \ref{section:FurtherReading} for a selection -- numerical evaluation of the likelihood function, $\pi(y_{obs}|\theta)$, is either computationally prohibitive, or simply not possible. Examples of the former can occur where the size of the observed dataset, $y_{obs}$, is sufficiently large that, in the absence of low dimensional sufficient statistics, evaluating the likelihood function even once is impracticable.  This can easily occur in the era of Big Data, for example, through large genomic datsets.
Partial likelihood intractability can arise, for instance, in models for Markov random fields. Here, the likelihood function can be written as
$p(y_{obs}|\theta) = \frac{1}{Z_{\theta}}\tilde{p}(y_{obs}|\theta)$
where $\tilde{p}(y_{obs}|\theta)$ is a function that can be evaluated, and where the normalisation constant, $Z_{\theta}=\sum_{\mathcal{Y}} \tilde{p}(y|\theta)$, depends on the parameter vector $\theta$.
Except for trivial datasets, the number of possible data configurations in the set $\mathcal{Y}$ means that brute-force enumeration of $Z_\theta$ is typically infeasible \shortcite{grelaud+rmrt09,moller+prb06}.
While there are algorithmic techniques available that arrange for the intractable normalising constants to cancel out within e.g. Metropolis-Hastings acceptance probabilities \shortcite{moller+prb06}, or that numerically approximate $Z_\theta$ through e.g. path sampling or thermodynamic integration, these are not viable when $\tilde{p}(y|\theta)$ itself is also computationally intractable.
Instances when the complete likelihood function is unavailable can also occur when the model density function is only implicitly defined, for example, through quantile or characteristic functions \shortcite{drovandi+p11,peters+sf12}. Similarly, the likelihood function may only be implicitly defined as a data generation process.

In these scenarios, if the preferred model is computationally intractable, the need to repeatedly evaluate the posterior distribution to draw samples from the posterior makes the implementation of standard Bayesian simulation techniques impractical.
Faced with this challenge,
one option is simply to fit a different model that is more amenable to statistical computations. The disadvantage of this approach is that the model could then be less realistic, and not permit inference on the particular questions of interest for the given analysis. A more attractive alternative, may be to consider an approximation to the preferred model, so that modelling realism is maintained at the expense of some approximation error. While various posterior approximation methods are available, ``likelihood-free'' Bayesian methods, of which approximate Bayesian computation (ABC) is a particular case, have emerged as an effective and intuitively accessible way of performing an approximate Bayesian analysis.

In this Chapter, we aim to give an intuitive exploration of the basics of ABC methods, illustrated wherever possible by simple examples. The scope of this exploration is deliberately limited, for example, we focus only on the use of simple rejection sampling based ABC samplers, in order that this Chapter will provide an accessible introduction to a subject which is given more detailed and advanced treatments in the rest of this Handbook.

\section{Likelihood-free intuition}%

The basic mechanism of likelihood-free methods can be fairly easily understood at an intuitive level.
For the moment, we assume that data generated under the model, $y\sim p(y|\theta)$, are discrete.
Consider the standard rejection sampling algorithm for sampling from a density $f(\theta)$:

\begin{table}[tbh]
\caption{\bf Standard Rejection Sampling Algorithm}
\noindent {\it Inputs:}
\begin{itemize}
\item A target density $f(\theta)$.
\item A sampling density $g(\theta)$, with $g(\theta)>0$ if $f(\theta)>0$.
\item An integer $N>0$.
\\
\end{itemize}

\noindent {\it Sampling:}\\
\noindent For $i=1, \ldots, N$:
\begin{enumerate}
\item \label{alg:rejection:step1} Generate $\theta^{(i)}\sim g(\theta)$ from sampling density $g$.
\item \label{alg:modified-rejection:step3}
Accept $\theta^{(i)}$ with probability  $\frac{f(\theta^{(i)})}{Kg(\theta^{(i)})}$ where $K\geq\max_\theta \frac{f(\theta)}{g(\theta)}$.\\
	Else go to \ref{alg:rejection:step1}.
\\
\end{enumerate}

\noindent {\it Output:}\\
A set of parameter vectors $\theta^{(1)},\ldots,\theta^{(N)}$ which are samples from $f(\theta)$.
\end{table}

If we specify $f(\theta)=\pi(\theta|y_{obs})$, and suppose that
the prior is used as the sampling distribution,
then the acceptance probability is proportional to the likelihood, as then $f(\theta)/Kg(\theta)\propto p(y_{obs}|\theta)$.  While direct evaluation of this acceptance probability is not available if the likelihood is computationally intractable, it is possible to stochastically determine whether or not to accept or reject a draw from the sampling density, {\it without} numerical evaluation of the acceptance probability. The following discussion assumes that the data $y$ are discrete (this will be relaxed later).

This can be achieved by noting that the acceptance probability is proportional to the probability of generating the observed data, $y_{obs}$,  under the model $p(y|\theta)$ for a fixed parameter vector, $\theta$. That is, suitably normalised, the likelihood function $p(y|\theta)$ can be considered as a probability mass function for the data. Put another way, for fixed $\theta$, if we generate a dataset from the model $y\sim p(y|\theta)$, then the probability of generating our observed dataset exactly, so that $y=y_{obs}$, is precisely $p(y_{obs}|\theta)$. From this observation, we can use the Bernoulli event of generating $y=y_{obs}$ (or not) to determine whether to accept (or reject) a draw from the sampling distribution, in lieu of directly evaluating the probability $p(y_{obs}|\theta)$.

This insight permits a rewriting of the simple rejection sampling algorithm, as given below.  A critical aspect of this modified algorithm is that it does not require numerical evaluation of the acceptance probability (i.e. the likelihood function). Note that if sampling is from $g(\theta)$ rather than the prior $\pi(\theta)$, then the acceptance probability is proportional to $p(y_{obs}|\theta)\pi(\theta)/g(\theta)$.  In this case, deciding whether to accept a draw from $g(\theta)$ can be split into two stages: firstly, as before, if we generate $y\sim p(y|\theta)$ such that $y\neq y_{obs}$ then we reject the draw from $g(\theta)$. If however, $y=y_{obs}$, then we accept the draw from $g(\theta)$ with probability 
$\pi(\theta)/[Kg(\theta)]$, where $K\geq\max_\theta f(\theta)/g(\theta)$. (These two steps may be interchanged so that the step with the least computational overheads is performed first.) Importance sampling versions of this and later algorithms are examined in \shortciteN{fan+s18}.

\begin{table}[tbh]
\caption{\bf Likelihood-Free Rejection Sampling Algorithm}
\noindent {\it Inputs:}
\begin{itemize}
\item A target posterior density $\pi(\theta|y_{obs})\propto p(y_{obs}|\theta)\pi(\theta)$, consisting of a prior distribution $\pi(\theta)$ and a procedure for generating data under the model  $p(y_{obs}|\theta)$.
\item A proposal density $g(\theta)$, with $g(\theta)>0$ if $\pi(\theta|y_{obs})>0$.
\item An integer $N>0$.
\\
\end{itemize}

\noindent {\it Sampling:}\\
\noindent For $i=1, \ldots, N$:
\begin{enumerate}
\item \label{alg:modified-rejection:step1} Generate $\theta^{(i)}\sim g(\theta)$ from sampling density $g$.
\item Generate $y\sim p(y|\theta^{(i)})$ from the likelihood.
\item \label{alg:modified-rejection:step3} If $y=y_{obs}$ then accept $\theta^{(i)}$ with probability  $\frac{\pi(\theta^{(i)})}{Kg(\theta^{(i)})}$,\\ where $K\geq\max_\theta\frac{\pi(\theta)}{g(\theta)}$.
	Else go to \ref{alg:modified-rejection:step1}.
\\
\end{enumerate}

\noindent {\it Output:}\\
A set of parameter vectors $\theta^{(1)},\ldots,\theta^{(N)}$ which are samples from $\pi(\theta|y_{obs})$.
\end{table}

\section{A practical illustration: Stereological extremes}%
\label{section:stereological}

In order to illustrate the performance of the likelihood-free rejection sampling algorithm, we perform a re-analysis of a stereological dataset with a computationally intractable model first developed by \shortciteN{bortot+cs07}.

\subsection{Background and model}

Interest is in the distribution of the size of {\em inclusions}, microscopically small particles introduced during the production of steel. The steel strength is thought to be directly related to the size of the largest inclusion. 
Commonly, the sampling of inclusions involves measuring the maximum cross-sectional diameter of each observed inclusion, $y_{obs}=(y_{obs,1}, \ldots, y_{obs,n})^\top$, obtained from a two-dimensional planar slice through the steel block. Each cross-sectional inclusion size is greater than some measurement threshold, $y_{obs,i}>u$.
The inferential problem is to analyse the unobserved distribution of the largest inclusion in the block, based on the information in the cross-sectional slice, $y_{obs}$. The focus on the size of the largest inclusion means that this is an extreme value variation on the standard stereological problem \cite{baddeley+j05}.

Each observed cross-sectional inclusion diameter, $y_{obs,i}$, is associated with an unobserved true inclusion diameter $V_i$.
\citeN{anderson+c02} proposed a mathematical model 
assuming that the inclusions were spherical with diameters $V$, and that their centres followed a homogeneous Poisson process with rate $\lambda>0$ in the volume of steel.
The distribution of the largest inclusion diameters, $V|V>v_0$
was assumed to follow a generalised Pareto distribution, with distribution function
\begin{equation}
\label{eqn:gpd}
	\mbox{Pr}(V\leq v|V>v_0) = 1-\left[1+\frac{\xi(v-v_0)}{\sigma}\right]^{-1/\xi}_+,
\end{equation}
for $v>v_0$, where $[a]_+=\max\{0,a\}$, 
following standard extreme value theory arguments \cite{coles01}. 
However, the probability of observing the cross-sectional diameter $y_{obs,i}$ (where $y_{obs,i}\leq V_i$) is dependent on the value of $V_i$, as larger inclusion diameters give a greater chance that the inclusion will be observed in the two-dimensional planar cross-section. This means that the number of observed inclusions, $n$, is also a random variable.
Accordingly the parameters of the full spherical inclusion model are  $\theta=(\lambda, \sigma, \xi)^\top$.

\citeN{anderson+c02} were able to construct a tractable likelihood function for this model by adapting the solution to Wicksell's corpuscle problem \cite{wicksell25}. 
However, while their model assumptions of a Poisson process 
are not unreasonable, the assumption that the inclusions are spherical is not plausible in practice.

\shortciteN{bortot+cs07} generalised this model to a family of ellipsoidal inclusions. 
While this model is more realistic than the spherical inclusion model, there are analytic and computational difficulties in extending likelihood-based inference to more general families of inclusion \shortcite{baddeley+j05,bortot+cs07}. As a result ABC methods are a good candidate procedure to approximate the posterior distribution in this case.

\subsection{Analysis}
\label{sec:extremesAnalysis}

For simplicity, suppose that we are interested in the spherical inclusions model, so that the true posterior distribution can be estimated directly. Suppose also that  the parameters of the generalised Pareto distribution are known to be $\sigma=1.5$ and $\xi=0.1$, so that interest is in the Poisson rate parameter, $\lambda$, only. In this setting, a sufficient statistic for the rate parameter is $n_{obs}$, the observed number of inclusions, so that $\pi(\theta|y_{obs})=\pi(\lambda|n_{obs})$ is the distribution of interest. 
Accordingly we can replace $y_{obs}=n_{obs}$ in the likelihood-free rejection sampling algorithm.
For the dataset considered by \shortciteN{bortot+cs07}, $n_{obs}=112$.

\begin{figure}[tb]
\centering
\includegraphics[width=12cm]{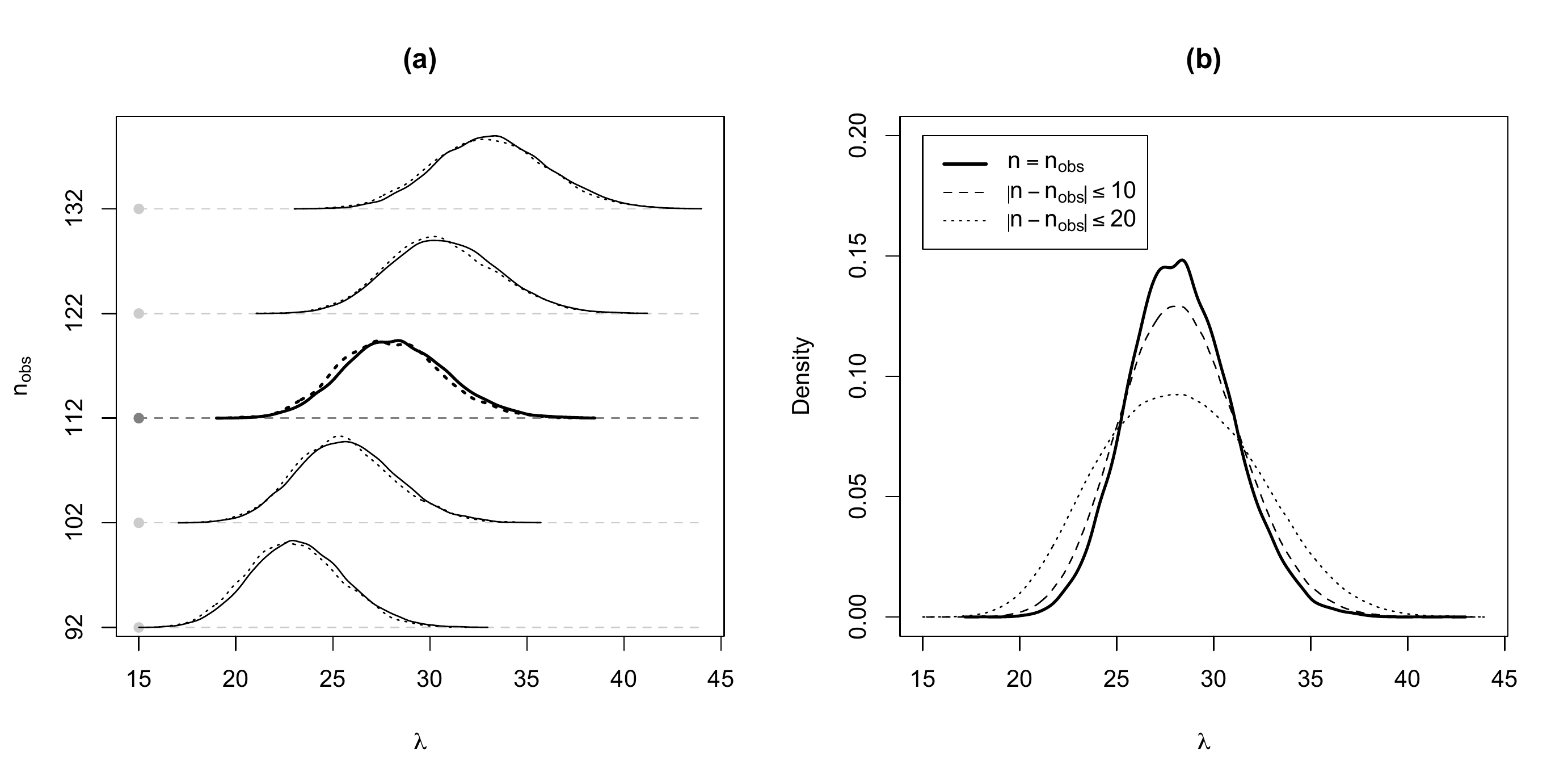}
\caption{\small Posterior density estimates of $\pi(\lambda|n_{obs})$ for the stereological extremes example, based on spherical inclusions.
(a) Density estimates using the likelihood-free rejection sampler (solid lines) and standard MCMC  algorithm (dashed lines), with $n_{obs}=92, 102, 112, 122$ and $132$.
(b) Density estimates for $n_{obs}=112$, with the relaxed criterion that $\|y-y_{obs}\|\leq h$ for $h=0, 10$ and $20$.
}
\label{chapter3:intro-bortot}
\end{figure}

Figure \ref{chapter3:intro-bortot}(a) shows scaled density estimates of $\pi(\lambda|n_{obs})$ (solid lines) obtained using the likelihood-free rejection sampling algorithm, for varying numbers of observed inclusions, $n_{obs}=92, 102, 112, 122$ and $132$. As the observed number of inclusions increases, accordingly so does the location and scale of the posterior of the rate parameter. The dashed lines in Figure \ref{chapter3:intro-bortot}(a) denote the same density estimates of $\pi(\lambda|n_{obs})$, but obtained using a conditional version of the standard MCMC sampler developed by \citeN{anderson+c02}, which makes use of numerical evaluations of the likelihood. These estimates are known to correspond to the true posterior. The likelihood-free rejection algorithm estimates clearly coincide with the true posterior distribution.

The density estimates obtained under the likelihood-free algorithm are each based on approximately 25,000 accepted samples, obtained from 5 million draws from the $U(0,100)$ prior. That is, the acceptance rate of the algorithm is approximately 0.5\%.
This algorithm is clearly very inefficient, with the computational overheads being partially influenced by the mismatch between prior and posterior distributions, but they are primarily dominated by the probability of generating data from the model that exactly matches the observed data, $n_{obs}$. This is the price for avoiding likelihood evaluation.
On balance, the computational inefficiency is practically acceptable for this specific case. However, this raises the question of how viable this approach will be for more complex analyses, when the probability of generating data such that $y=y_{obs}$ becomes even lower. Further, the acceptance probability will be exactly zero if the data generated under the model, $y\sim p(y|\theta)$, are continuous, which is likely to be the case in general.

In order to alleviate such computational overheads,
one possible variation on the likelihood-free rejection algorithm would be to adjust the potentially very low (or zero) probability requirement that $y=y_{obs}$ exactly. Instead,  the acceptance criterion could require that the generated data is simply ``close'' to the observed data. For example, this might require that $\|y-y_{obs}\|\leq h$ for some $h\geq 0$ and distance measure $\|\cdot\|$, such as Euclidean distance. This would also permit a relaxation of our previous assumption that data generated under the model, $y\sim p(y|\theta)$, are discrete. In this way, step \ref{alg:modified-rejection:step3} of the {\it Sampling} stage of the likelihood-free rejection algorithm would become:

\begin{table}
\caption{\bf Likelihood-Free Rejection Sampling Algorithm}

\begin{enumerate}
\item[3.] If $\|y-y_{obs}\|\leq h$ then accept $\theta^{(i)}$ with probability  $\frac{\pi(\theta^{(i)})}{Kg(\theta^{(i)})}$,\\ where $K\geq\max_\theta\frac{\pi(\theta)}{g(\theta)}$.

	Else go to \ref{alg:modified-rejection:step1}.
\end{enumerate}

\end{table}

Of course, the output samples would no longer be draws from $\pi(\theta|y_{obs})$ unless $h=0$, but will instead be draws from an approximation of $\pi(\theta|y_{obs})$.

The logic behind this modification is that increasing $h$ will considerably improve the acceptance rate of the algorithm. The hope is that, if $h$ remains small, then the resulting estimate of the posterior will still be close to the true posterior. An illustration of this is shown in Figure \ref{chapter3:intro-bortot}(b), which shows density estimates obtained using the adjusted requirement that $\|n-n_{obs}\|\leq h$ for $h=0$ (i.e. $n=n_{obs}$), $10$ and $20$. Computationally there is a marked improvement in algorithmic efficiency: the low $0.5\%$ acceptance rate for $h=0$ increases to $10.5\%$ and $20.5\%$ for $h=10$ and $20$ respectively.

However, there are now some clear deviations in the density estimate resulting from the likelihood-free algorithm, compared to the actual posterior, $\pi(\lambda|n_{obs})$ (solid lines). In fact, it is more accurate to refer to these density estimates as an approximation of the posterior. On one hand, the location and shape of the density are broadly correct, and for some applications, this level of approximation may be adequate. On the other hand, however, the scale of the approximation is clearly overestimated for larger values of $h$. Intuitively this makes sense: the adjusted criterion $\|y-y_{obs}\|\leq h$ accepts $\theta\sim g(\theta)$ draws if the generated data $y$ is merely ``close'' to $y_{obs}$. As such, for many values of $\theta$ where it was previously very unlikely to generate data such that $y=y_{obs}$, it may now be possible to satisfy the more relaxed criterion. This will accordingly result in a greater range of $\theta$ values that will be accepted, and thereby increase the variability of the posterior approximation. The more relaxed the criterion (i.e. the larger the value of $h$), the greater the resulting variability.

It is possible to be more precise about the exact form of the posterior obtained through this adjusted procedure -- this will be discussed in detail in the next Section. However, for this particular analysis, based on samples $\lambda^{(1)},\ldots,\lambda^{(N)}$ and datasets $n^{(1)},\ldots,n^{(N)}$ obtained from the likelihood-free rejection algorithm, it can be seen that as the posterior approximation is constructed from those values of $\lambda=\lambda^{(i)}$ such that $\|n^{(i)}-n_{obs}\|\leq h$, then the posterior approximation can firstly be expressed as
\begin{eqnarray*}
	\hat{\pi}(\lambda|n_{obs})  = \frac{1}{N} \sum_{i=1}^N\delta_{\lambda^{(i)}}(\lambda)
	&=&
	\frac{1}{N}\sum_{\lambda^{(i)}:\|n^{(i)}-n_{obs}\|\leq h}\delta_{\lambda^{(i)}}(\lambda)\\
	& = &
	 \sum_{h'=-h^*}^{h^*}\left(\frac{1}{N}\sum_{\lambda^{(i)}:(n^{(i)}-n_{obs})=h'}\delta_{\lambda^{(i)}}(\lambda)\right),
\end{eqnarray*}
where $h^*$ is the largest integer such that $\|h^*\|\leq h$.
It then follows that
\begin{equation}
\label{eqn:discreteMixturePost}
	\lim_{N\rightarrow\infty} \hat{\pi}(\lambda|n_{obs})
	 =
	\sum_{h'=-h^*}^{h^*} \mbox{Pr}(n=n_{obs}+h')\pi(\lambda|n_{obs}+h').
\end{equation}
That is, the ``likelihood-free'' approximation  of the posterior, $\pi(\theta|y_{obs})$, is precisely an average of the individual posterior distributions $\pi(\lambda|n_{obs}+h')$ for $h'=-h^*,\ldots,h^*$, weighted according to $\mbox{Pr}(n=n_{obs}+h')$, the probability of observing the dataset, $n_{obs}+h'$, based on samples drawn from the (prior predictive) distribution $p(n|\lambda)\pi(\lambda)$.
This can be loosely observed from Figure \ref{chapter3:intro-bortot}, in which the approximations for $h=10$  and $h=20$ in panel (b) respectively correspond to rough visual averages of the centre three and all five displayed posteriors in panel (a). For $h=0$ we obtain $\lim_{N\rightarrow\infty}\hat{\pi}(\lambda|n_{obs})=\pi(\lambda|n_{obs})$ as for standard Monte Carlo algorithms.

Similar interpretations and conclusions arise when the data $y$ are continuous, as we examine for a different model in the following Subsection. This also allows us to introduce a fundamental concept in ABC methods -- the use of summary statistics.

\section{A $g$-and-$k$ distribution analysis}%
\label{section:gandk}

The univariate $g$-and-$k$ distribution is a flexible unimodal distribution that is able to describe data with significant amounts of skewness and kurtosis.
Originally developed by 
\citeN{tukey77} (see also \citeNP{martinez+i84,hoaglin85} and
\citeNP{rayner+m02}), the $g$-and-$k$ and related distributions have been analysed in the ABC setting by \shortciteN{peters+s06}, \shortciteN{allingham+km09}, \shortciteN{drovandi+p11} and \shortciteN{fearnhead+p12} among others. Its density function has no closed form, but is alternatively defined through its quantile function as
\begin{eqnarray}
\label{eqn:g&k}
	Q(q|A,B,g,k) = A + B\left[1+c\frac{1-\exp\{-gz(q)\}}{1+\exp\{-gz(q)\}}\right] (1+z(q)^2)^k z(q)
\end{eqnarray}
for $B>0, k>-1/2$, 
where $z(q)=\Phi^{-1}(q)$ is the $q$-th quantile of the standard normal distribution function. The parameter $c$ measures overall asymmetry, and is conventionally fixed at $c=0.8$  (resulting in $k>-1/2$) \cite{rayner+m02}. This distribution is very flexible, with many common distributions obtained or well approximated by particular parameter settings, such as the normal distribution when $g=k=0$. Given $\theta=(A,B,g,k)^\top$, simulations $z(q)\sim N(0,1)$ drawn from a standard normal distribution can be transformed into samples from the $g$-and-$k$ distribution through equation (\ref{eqn:g&k}).
	
\begin{figure}[tb]
\centering
\includegraphics[width=12cm]{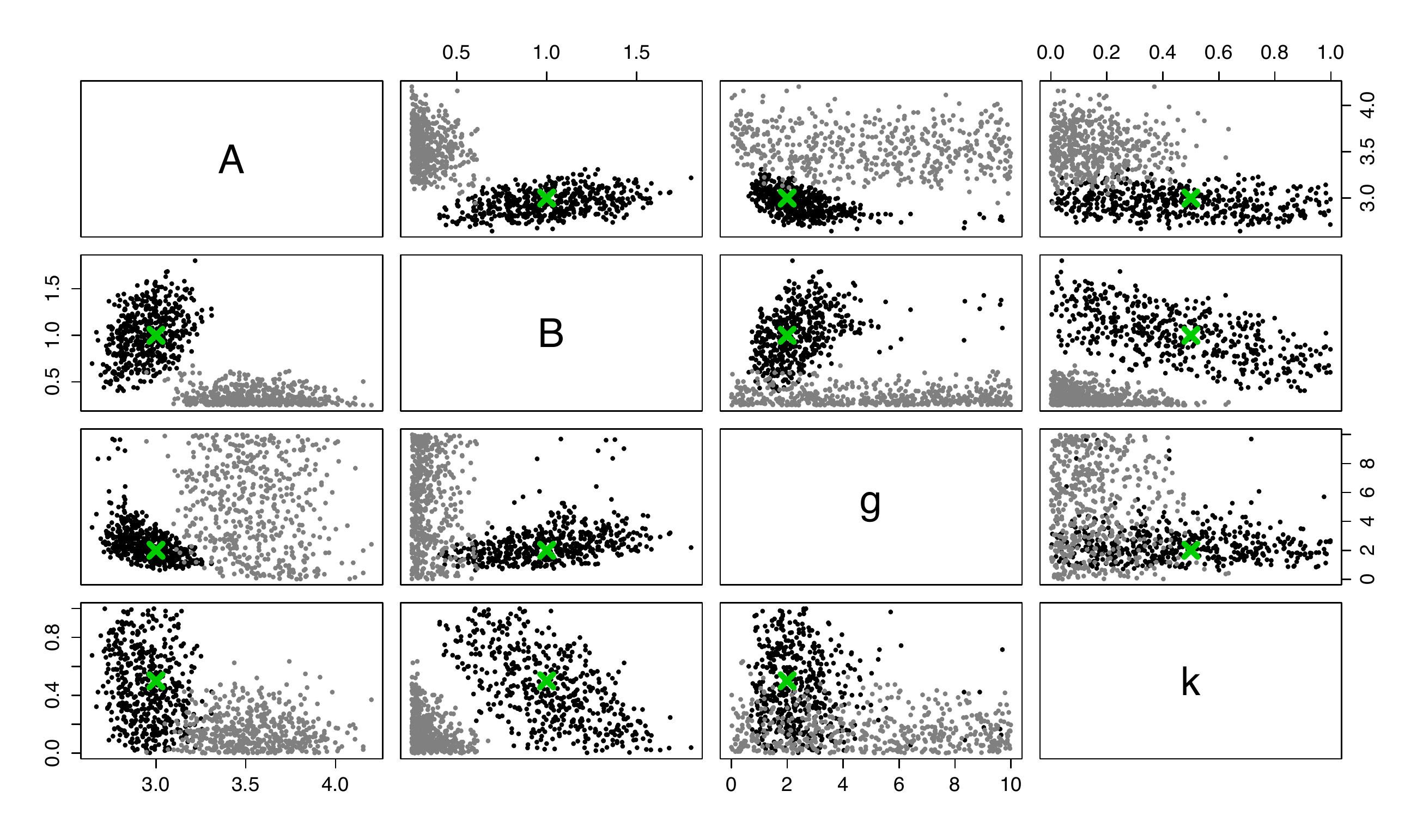}
\caption{{\protect\small Pairwise scatterplots of samples from the likelihood-free approximation to the posterior using the full dataset (grey dots), and four summary statistics (black dots). True parameter values $(A,B,g,k)=(3,1,2,0.5)$ are indicated by the cross $\times$.
}}
\label{image:gandk}
\end{figure}

Figure \ref{image:gandk} shows a scatterplot of samples from the likelihood-free approximation of the posterior $\pi(\theta|y_{obs})$ (grey dots), based on a simulated dataset $y_{obs}$ of length $n=1,000$ generated from the $g$-and-$k$ distribution with parameter vector $\theta_0=(3,1,2,0.5)^\top$. 
This analysis was based on defining 
$\|y-y_{obs}\| = (y-y_{obs})^\top\hat{\Sigma}^{-1}(y-y_{obs})\leq h$ as Mahalanobis distance, with $h$ given by the 0.005 quantile of the differences $\|y-y_{obs}\|$ for $i=1,\ldots, N=100,000$ Monte Carlo samples from the joint prior $\pi(\theta)=\pi(A)\pi(B)\pi(g)\pi(k)=N(1,5)\times N(0.25,2)\times U(0,10)\times U(0,1).$
The matrix $\hat{\Sigma}$ was determined as the sample covariance matrix of $y$ using 2,000 samples generated under the model $y|\theta_0$ with $\theta=\theta_0$ fixed at its true value.

As is apparent from Figure \ref{image:gandk}, the likelihood-free approximation to $\pi(\theta|y_{obs})$ (grey dots) is particularly poor -- the true parameter vector $\theta_0$ is not even close to  the estimated posterior samples. This outcome is a direct result of the dimension of the comparison $y-y_{obs}$. The chance of generating an $n=1,000$-dimensional vector $y$ that is close to $y_{obs}$, even if $\theta=\theta_0$, is vanishingly small. 
The odds of matching $y$ with $y_{obs}$ can be increased by redefining both in terms of their order statistics, although the chances still remain extremely low (see Example 3 in Section \ref{section:summaryStatisticBasics} for an illustration).
This means that $h$ must be relatively large, which results in accepting samples $\theta^{(i)}$ that generate data $y^{(i)}$ that are not actually close to $y_{obs}$, and thereby producing a poor approximation to $\pi(\theta|y_{obs})$.

The obvious way to avoid this problem is to reduce the dimension of the data comparison $y-y_{obs}$. Suppose that lower dimensional statistics $s=S(y)$ and $s_{obs}=S(y_{obs})$ are available, such that $S(y)$ is sufficient for, or highly informative for $\theta$ under the model, but where $\dim(S(y))\ll\dim(y)$. Then the comparison $\|y-y_{obs}\|$ might be replaced by $\|s-s_{obs}\|$ without too much loss of information, but with the advantage that the dimension of $S(y)$ is now much lower. That is, step 3 in the likelihood-free rejection sampling algorithm could be further replaced by:

\begin{table}
\caption{\bf Likelihood-Free Rejection Sampling Algorithm}

\begin{enumerate}
\item[3.] Compute $s=S(y)$.\\
If $\|s-s_{obs}\|\leq h$ then accept $\theta^{(i)}$ with probability  $\frac{\pi(\theta^{(i)})}{Kg(\theta^{(i)})}$\\ where $K\geq\max_\theta \frac{\pi(\theta)}{g(\theta)}$.
	Else go to \ref{alg:modified-rejection:step1}.
\end{enumerate}

\end{table}

Using this idea, \shortciteN{drovandi+p11} suggested the statistics
\begin{eqnarray*}
	S_A&=&E_4,
	\quad
	S_B=E_6-E_2,
	\quad
	S_g=(E_6+E_2-2E_4)/S_B,\\
	\mbox{and }
	S_k&=&(E_7-E_5+E_3-E_1)/S_B
\end{eqnarray*}
as informative for $A, B, g$ and $k$ respectively, so that $S(y)=(S_A,S_B,S_g,S_k)^\top$, where
$E_1\leq E_2\leq\ldots\leq E_8$ are the octiles of $y$. Repeating the above $g$-and-$k$ analysis but using the 4-dimensional comparison $\|s-s_{obs}\|$ rather than $\|y-y_{obs}\|$ (and recomputing $\hat{\Sigma}$ and $h$ under the same conditions), the resulting posterior samples are shown in Figure \ref{image:gandk} (black dots).

The difference in the quality of the approximation to $\pi(\theta|y_{obs})$ when using $S(y)$ rather than $y$, is immediately apparent. The true parameter value $\theta_0$ is now located firmly in the centre of each pairwise posterior sample, several parameters (particularly $A$ and $g$) are more precisely estimated, and evidence of dependence between parameters (as is to be expected) is now
clearly seen. 

While it is unreasonable to expect that there has been no loss of information in moving from $y$ to $S(y)$, clearly the overall gain in the quality of the approximation to the likelihood-free posterior
has been worth it in this case. This suggests that the use of summary statistics $S(y)$ is a useful tool more generally in approximate Bayesian computational techniques.

\section{Likelihood-free methods or approximate Bayesian computation (ABC)?}%

The terms {\em likelihood-free} methods and {\em approximate Bayesian computation} are both commonly used to describe Bayesian computational methods developed for when the likelihood function is computationally intractable, or otherwise unavailable. Of course, ``likelihood-free'' is arguably a misnomer -- in no sense is the likelihood function not involved in the analysis. It is the function used to generate the data $y\sim p(y|\theta)$, and it accordingly must exist, whether or not it can be numerically evaluated or written down. Rather, in this context, ``likelihood-free'' refers to any likelihood-based analysis that proceeds without direct numerical evaluation of the likelihood function. There are several techniques that could be classified according to this description.

``Approximate Bayesian computation'', commonly abbreviated to ``ABC'', was first coined by  \shortciteN{beaumont+zb02} in the context of Bayesian statistical techniques in population genetics (although see \citeNP{tavare17}, this volume), and refers to the specific type of likelihood-free methods considered in this book. In particular, given the ``approximate'' in ABC, it refers to those likelihood-free methods that produce an approximation to the posterior distribution resulting from the imperfect matching of data $\|y-y_{obs}\|$ or summary statistics $\|s-s_{obs}\|$.

Thus, the likelihood-free rejection algorithm described above with $h=0$, which only accepts samples,  $\theta$, which have exactly reproduced the observed data $y_{obs}$, is not an ABC algorithm, as the method produces exact samples from the posterior distribution -- there is no approximation. (The Monte Carlo approximation of the posterior is not considered an approximation in this sense.) It is, however, a likelihood-free method.  Whereas, the likelihood-free rejection algorithm which may accept samples if $\|y-y_{obs}\|\leq h$, for $h>0$, is an ABC algorithm, as the samples will be drawn from an approximation to the posterior distribution.
Similarly, when the sampler may alternatively accept samples if $\|s-s_{obs}\|\leq h$, for any $h\geq 0$ (including $h=0$), the resulting samples are also drawn from an approximate posterior distribution. As such, this is also an ABC algorithm. The only exception to this is the case where $h=0$ and the summary statistics are sufficient: here there is no posterior approximation -- the algorithm is then likelihood-free but not an ABC method.  

With a few exceptions (such as indirect inference, see \shortciteNP{drovandi18}) all of the methods considered in this book are both ABC  and (by definition) likelihood-free methods. The aim of any ABC analysis is to find a practical way of performing the Bayesian analysis, while keeping the Approximation and the Computation to a minimum.

\section{The approximate posterior distribution}
\label{chapter3:section:TheApproximatePosteriorDistribution}

In contrast to the intuitive development of likelihood-free methods in the previous Sections,
we now describe the exact form of the ABC approximation to the posterior distribution that is produced from the  likelihood-free rejection algorithm.
The procedure of (i) generating $\theta$ from the sampling distribution, $g(\theta)$,  (ii) generating data, $y$, from the likelihood, $p(y|\theta)$, conditional on $\theta$, and (iii) rejecting $\theta$ if $\|y-y_{obs}\|\leq h$,
is equivalent to drawing a sample $(\theta,y)$ from the joint distribution proportional to
\[
I(\|y-y_{obs}\leq h\|)p(y|\theta)g(\theta),
\]
where $I$ is the indicator function, with  $I(Z)=1$ if $Z$ is true, and $I(Z)=0$ otherwise.
If this sample $(\theta,y)$ is then further accepted with probability proportional to $\pi(\theta)/g(\theta)$, this implies that the likelihood-free rejection algorithm is sampling from the joint distribution proportional to
\begin{equation}
\label{eqn:simplejoint}
	I(\|y-y_{obs}\|\leq h)p(y|\theta)g(\theta)\frac{\pi(\theta)}{g(\theta)}\\
	=
	I(\|y-y_{obs}\|\leq h)p(y|\theta)\pi(\theta).
\end{equation}
Note that if $h=0$, then the $\theta$ marginal of (\ref{eqn:simplejoint}) equals the true posterior distribution, as
\begin{eqnarray*}
	\lim_{h\rightarrow 0}\int I(\|y-y_{obs}\|\leq h)p(y|\theta)\pi(\theta) dy
	& = &
	\int\delta_{y_{obs}}(y)p(y|\theta)\pi(\theta) dy\\
	& = &
	p(y_{obs}|\theta)\pi(\theta).
\end{eqnarray*}
That is, for $h=0$, the likelihood-free rejection algorithm draws samples, $(\theta,y)$, for which the marginal distribution of the parameter vector is the true posterior, $\pi(\theta|y_{obs})$. (The marginal distribution of the auxiliary dataset $y$ is a point mass at $\{y=y_{obs}\}$ in this case.)

It is useful in the following to generalise the above formulation slightly. In (\ref{eqn:simplejoint}), the indicator  term $I(\|y-y_{obs}\leq h\|)$ only takes the values 0 or 1. This is useful in the sense that it allows clear ``{\it If $\|y-y_{obs}\|\leq h$ then \ldots }'' statements to be made in any   algorithm, which can simplify implementation. However it is intuitively wasteful of information, as it does not discriminate between those samples, $\theta$, for which the associated dataset $y$ exactly equals the observed dataset $y_{obs}$, and those samples, $\theta$, for which the associated dataset is the furthest away from $y_{obs}$, i.e. $\|y-y_{obs}\|=h$. As the former case produces samples that are exact draws from the true posterior distribution, whereas the latter case does not, this produces a motivation for a more continuous scaling from 1 (when $y=y_{obs}$) to 0 (when $\|y-y_{obs}\|$ is large).

\begin{figure}[tbh!]
\centering
\includegraphics[width=9cm]{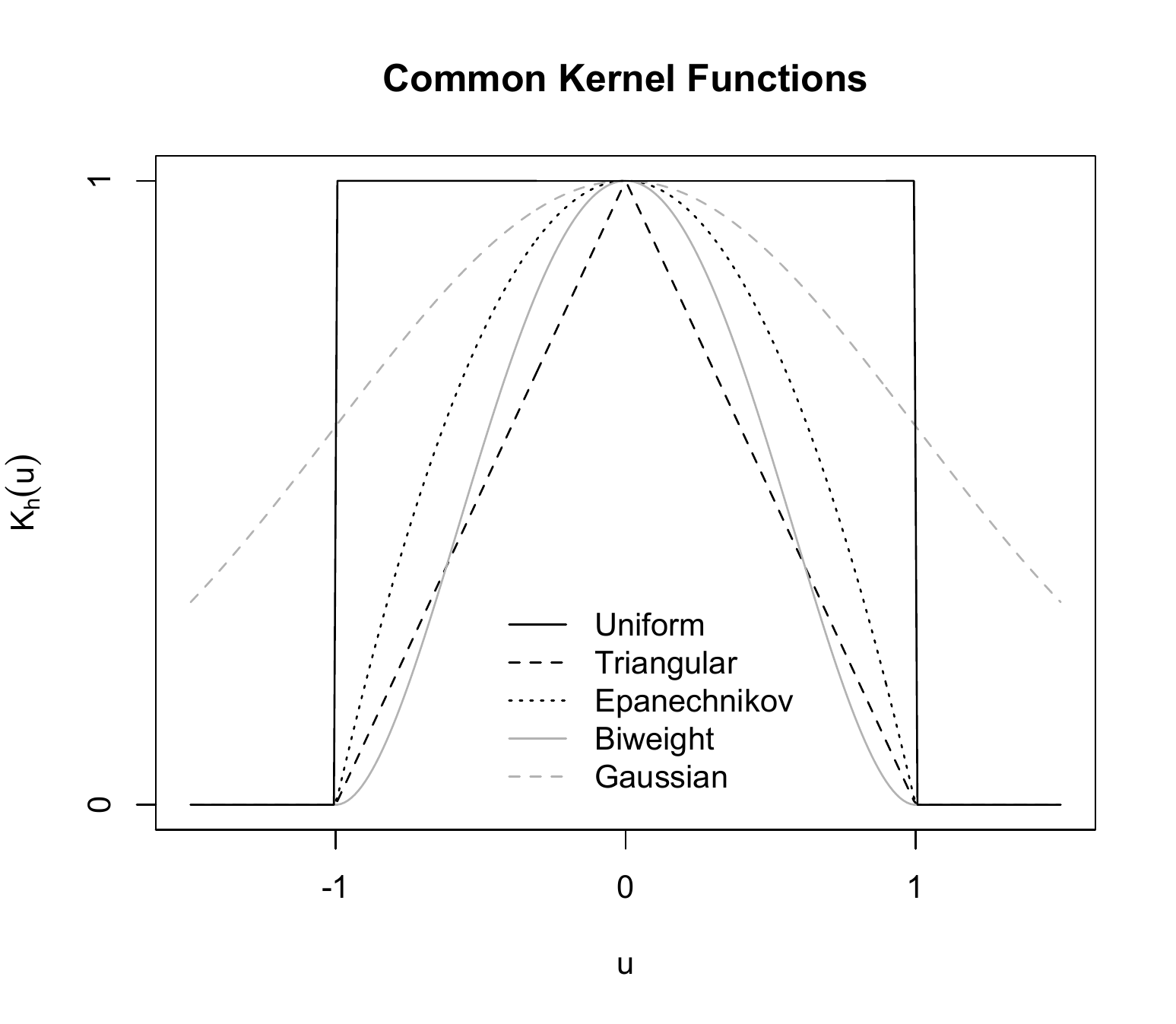}
\caption{\small Standard  kernel functions, $K(u)$, listed in Table \ref{Chapter3:table:StandardKernels} plotted on a common scale (with maximum at $1$).
}
\label{chapter3:figure:StandardKernels}
\end{figure}

This can be achieved by replacing the indicator function, $I(\|y-y_{obs}\|\leq h)$, with a standard smoothing kernel function, $K_h(u)$, with $u=\|y-y_{obs}\|$, where
\[
K_h(u)=\frac{1}{h}K\left(\frac{u}{h}\right).
\]
Kernels are symmetric functions such that $K(u)\geq 0$ for all $u$, $\int K(u)du=1$, $\int uK(u)du=0$ and $\int u^2K(u)du<\infty$.
Here, $h>0$ corresponds to the scale parameter, or ``bandwidth'' of the kernel function. Several common forms for kernel functions are given in Table \ref{Chapter3:table:StandardKernels}, and these are illustrated in Figure \ref{chapter3:figure:StandardKernels}. Following convention, we define $\lim_{h\rightarrow 0}K_h(u)$ as a point mass at the origin ($u=0$).

\begin{table}[tbh!]
\caption{\small The functional forms of several common kernel functions.}
\label{Chapter3:table:StandardKernels}
\setlength{\tabcolsep}{0.25 cm}
\begin{center}
\begin{tabular}{ll}
Kernel & $K(u)$  \\ \hline
\vspace{1mm}
Uniform & $\frac{1}{2}I(|u|\leq 1)$\\
\vspace{1mm}
Triangular & $(1-|u|)I(|u|\leq 1)$\\
\vspace{1mm}
Epanechnikov & $\frac{3}{4}(1-u^2)I(|u|\leq 1)$  \\
\vspace{1mm}
Biweight & $\frac{15}{16}(1-u^2)^3I(|u|\leq 1)$ \\
Gaussian & $\frac{1}{\sqrt{2\pi}}e^{-\frac{1}{2}u^2}$\\
\end{tabular}
\end{center}
\end{table}

An alternative specification of a smoothing kernel for multivariate datasets is obtained by writing $u=y-y_{obs}$, where $u=(u_1,\ldots,u_n)^\top$, $y=(y_1,\ldots,y_n)^\top$ and $y_{obs}=(y_{obs,1},\ldots,y_{obs,n})^\top$, so that $u_i=y_i-y_{obs,i}$.  Then we can write
$K_h(u)=\prod_{i=1}^n K_{h_i}(u_i)$, where the scale parameter of each individual kernel function, $K_{h_i}(u_i)$, may vary. A further, more general specification may  determine $K_h(u)$ as a fully multivariate, smooth and symmetric function, satisfying the above moment constraints. One such example is a multivariate $N(0,\Sigma)$ distribution, for some fixed covariance matrix $\Sigma$.

Substituting
the kernel function, $K_h(u)$, into the likelihood-free rejection algorithm results in the ABC Rejection Sampling Algorithm:

\begin{table}
\caption{\bf ABC Rejection Sampling Algorithm}
\noindent {\it Inputs:}
\begin{itemize}
\item A target posterior density $\pi(\theta|y_{obs})\propto p(y_{obs}|\theta)\pi(\theta)$, consisting of a prior distribution $\pi(\theta)$ and a procedure for generating data under the model  $p(y_{obs}|\theta)$.
\item A proposal density $g(\theta)$, with $g(\theta)>0$ if $\pi(\theta|y_{obs})>0$.
\item An integer $N>0$.
\item A kernel function $K_h(u)$ and scale parameter $h>0$.
\\
\end{itemize}

\noindent {\it Sampling:}\\
\noindent For $i=1, \ldots, N$:
\begin{enumerate}
\item \label{chapter3:alg:ABC-rejection:step1} Generate $\theta^{(i)}\sim g(\theta)$ from sampling density $g$.
\item Generate $y\sim p(y|\theta^{(i)})$ from the likelihood.
\item Accept $\theta^{(i)}$ with probability $\frac{K_h(\|y-y_{obs}\|)\pi(\theta^{(i)})}{Kg(\theta^{(i)})}$\\ where $K\geq K_h(0)\max_\theta \frac{\pi(\theta)}{g(\theta)}$.
	Else go to \ref{chapter3:alg:ABC-rejection:step1}.
\\
\end{enumerate}

\noindent {\it Output:}\\
A set of parameter vectors $\theta^{(1)},\ldots,\theta^{(N)}$ $\sim$ $\pi_{ABC}(\theta|y_{obs})$.
\end{table}

In order to determine the form of the target distribution, $\pi_{ABC}(\theta|y_{obs})$, of this algorithm, we can
follow the same argument as before. By (i) generating $\theta$ from the importance distribution, $g(\theta)$, (ii) generating data, $y$, from the likelihood, $p(y|\theta)$, conditional on $\theta$, and then (iii) accepting the sample $(\theta,y)$ with probability proportional to $K_h(\|y-y_{obs}\|)\pi(\theta^{(i)})/g(\theta^{(i)})$, this results in samples from the joint distribution
\begin{equation}
\label{Chapter3:eqn:ABCjointposterior}
	\pi_{ABC}(\theta,y|y_{obs}) \propto K_h(\|y-y_{obs}\|)p(y|\theta)\pi(\theta).
\end{equation}
When $K_h(u)$ is the uniform kernel (see Table \ref {Chapter3:table:StandardKernels}), then (\ref{Chapter3:eqn:ABCjointposterior}) reduces to (\ref{eqn:simplejoint}).
Accordingly, we define the ABC approximation to the true posterior distribution as
\begin{equation}
\label{Chapter3:eqn:ABCposterior}
	\pi_{ABC}(\theta|y_{obs}) = \int \pi_{ABC}(\theta,y|y_{obs}) dy,
\end{equation}
where $\pi_{ABC}(\theta,y|y_{obs})$ is given by (\ref{Chapter3:eqn:ABCjointposterior}).

As before, as $h\rightarrow 0$, so that only those samples, $\theta$, that generate data for which $y=y_{obs}$ are retained,
then (\ref{Chapter3:eqn:ABCjointposterior}) becomes
\begin{eqnarray*}
	\lim_{h\rightarrow 0} \pi_{ABC}(\theta,y|y_{obs})
	& \propto & \lim_{h\rightarrow 0} K_h(\|y-y_{obs}\|)p(y|\theta)\pi(\theta)\\
	& = & \delta_{y_{obs}}(y)p(y|\theta)\pi(\theta),
\end{eqnarray*}
and so 
	$\lim_{h\rightarrow 0}\pi_{ABC}(\theta|y_{obs})\propto\int\delta_{y_{obs}}(y)p(y|\theta)\pi(\theta)dy
	=p(y_{obs}|\theta)\pi(\theta).$
That is, samples from the true posterior distribution are obtained as $h\rightarrow 0$. However, $h=0$ is not a viable choice in practice, as for continuous $y_{obs}$ it corresponds to an algorithm with an acceptance rate of zero. 

To see what marginal distribution the ABC rejection algorithm is sampling from for $h>0$ we can integrate $\pi_{ABC}(\theta,y|y_{obs})$ over the  auxiliary data margin, $y$.

A natural question to ask is, how accurate is this approximation? Re-writing the right hand side of (\ref{Chapter3:eqn:ABCjointposterior}) without the prior distribution, $\pi(\theta)$, we can similarly define the ABC approximation to the true likelihood, $p(y|\theta)$, for a fixed value of $\theta$, as
\begin{equation}
\label{Chapter3:eqn:ABClikelihood}
	p_{ABC}(y_{obs}|\theta) = \int K_h(\|y-y_{obs}\|)p(y|\theta) dy.
\end{equation}
In this manner, ABC can be interpreted as a regular Bayesian analysis, but with an approximated likelihood function.

 Working in the univariate case for simplicity of illustration, so that $y,y_{obs}\in{\mathcal Y}=\mathbb{R}$ and $\|u\|=|u|$, we can obtain
\begin{eqnarray}
	p_{ABC}(y_{obs}|\theta) & = & \int K_h(|y-y_{obs}|)p(y|\theta)dy\nonumber \\
	& = &
	\int K(u)p(y_{obs}-uh|\theta)du\nonumber\\
	& = &
	\int K(u)\left[p(y_{obs}|\theta) - uhp'(y_{obs}|\theta) + \frac{u^2h^2}{2}p''(y_{obs}|\theta) - \ldots \right] du\nonumber\\
	& = &
	p(y_{obs}|\theta)+ \frac{1}{2}h^2p''(y_{obs}|\theta)\int u^2K(u)du - \ldots\label{eqn:conDenEst}
\end{eqnarray}
using the substitution $u=(y_{obs}-y)/h$, a Taylor expansion of $p(y_{obs}-uh|\theta)$ around the point $y_{obs}$, and the kernel function properties of $K_h(u)=K(u/h)/h$, $\int K(u)du=1$, $\int uK(u)du=0$ and $K(u)=K(-u)$.
The above is a standard smoothing kernel density estimation expansion, and assumes that the likelihood, $p(y|\theta)$, is infinitely differentiable. As with kernel density estimation, the choice of scale parameter is more important than the choice of kernel function in terms of the quality of the approximation.

Then,  the pointwise bias in the likelihood approximation for fixed $\theta$ can be expressed as
\begin{equation}
\label{Chapter3:eqn:b}
	b_h(y|\theta) := p_{ABC}(y|\theta)-p(y|\theta),
\end{equation}
as a function of $y$,
which to second order can be written as
\begin{equation*}
	\hat{b}_h(y|\theta) =  \frac{1}{2}h^2\sigma^2_Kp''(y|\theta),
\end{equation*}
where $\sigma^2_K=\int u^2K(u)du$ is the variance of the kernel function. Accordingly,
the magnitude of the bias is reduced if $h$ is small, corresponding to better approximations. Clearly, the second derivative of the likelihood function, $p''(y|\theta)$, is typically also unavailable if the likelihood function itself is computationally intractable.
When $y,y_{obs}\in{\mathcal Y}$ is multivariate, a similar derivation to the above is available. In either case, the ABC approximation to the true posterior is defined through (\ref{Chapter3:eqn:ABCposterior}).

In a similar manner, we can determine the pointwise bias in the resulting ABC posterior approximation.
From (\ref{Chapter3:eqn:b}) we can write
\begin{eqnarray}
\label{Chapter3:eqn:bit}
	b_h(y_{obs}|\theta)\pi(\theta) &  = & p_{ABC}(y_{obs}|\theta)\pi(\theta)-p(y_{obs}|\theta)\pi(\theta)\nonumber\\
	& = & \pi_{ABC}(\theta|y_{obs})c_{ABC}-\pi(\theta|y_{obs})c,
\end{eqnarray}
where $c_{ABC}=\int p_{ABC}(y_{obs}|\theta)\pi(\theta)d\theta>0$ and $c=\int p(y_{obs}|\theta)  \pi(\theta)d\theta>0$. Rearranging (\ref{Chapter3:eqn:bit}), we obtain
\begin{eqnarray}
\label{eqn:ahat}
	a_h(\theta|y_{obs}) & := &
	\pi_{ABC}(\theta|y_{obs})-\pi(\theta|y_{obs})\nonumber\\
	& \textcolor{red}{=} & \frac{b_h(y_{obs}|\theta)\pi(\theta) + \pi(\theta|y_{obs})c}{c_{ABC}}-\pi(\theta|y_{obs})\\
	& = & \frac{b_h(y_{obs}|\theta)\pi(\theta)}{c_{ABC}}+\pi(\theta|y_{obs})\left(\frac{c}{c_{ABC}}-1\right),\nonumber
\end{eqnarray}
as a function of $\theta$. As $h\rightarrow 0$, then $b_h(y_{obs}|\theta)\rightarrow 0$ from (\ref{Chapter3:eqn:b}), and so $p_{ABC}(y_{obs}|\theta)\rightarrow p(y_{obs}|\theta)$ pointwise, for fixed $\theta$. Further, $c/c_{ABC}\rightarrow 1$ as $h$ gets small, so that $a_h(\theta|y_{obs})\rightarrow 0$.

\subsection{Simple examples}%

In many simple cases, the ABC approximation to the posterior distribution can be derived exactly.
\\

\noindent {\bf Example 1:}\\
 \noindent Suppose that the observed data, $y_{obs}$, is a single draw from a univariate density function $p(y|\theta)$, and that $\theta$ is a scalar. If we consider the particular case where $K_h(\|u\|)$ is the uniform kernel on $[-h,h]$ (see Table \ref{Chapter3:table:StandardKernels}), and $\|u\|=|u|$, then we have
 \begin{eqnarray}
 \pi_{ABC}(\theta|y_{obs}) &\propto & \pi(\theta)\int_{-\infty}^{\infty} K_h(|y-y_{obs}|)p(y|\theta)dy\nonumber\\
& = &
 \frac{\pi(\theta)}{2h}\int_{y_{obs}-h}^{y_{obs}+h}p(y|\theta)dy\nonumber\\
& = &
 \pi(\theta)\frac{\left[P(y_{obs}+h|\theta)-P(y_{obs}-h|\theta)\right]}{2h}, \label{easyex1}
 \end{eqnarray}
where $P(y|\theta)=\int_{-\infty}^y p(z|\theta)dz$ is the cumulative distribution function of $y|\theta$. Noting that as
$\lim_{h\rightarrow 0}[P(y_{obs}+h|\theta)-P(y_{obs}-h|\theta)]/2h = p(y_{obs}|\theta)$
via l'Hopital's rule,  then $\pi_{ABC}(\theta|y_{obs})\rightarrow \pi(\theta|y_{obs})$ as $h\rightarrow 0$, as required. Also,
$[P(y_{obs}+h|\theta)-P(y_{obs}-h|\theta)]/2h\approx 1/2h$ for large $h$, and so $\pi_{ABC}(\theta|y_{obs})\rightarrow \pi(\theta)$ as $h\rightarrow\infty$.

Suppose now that $p(y|\theta)=\theta e^{-\theta y}$, for $\theta, y\geq 0$, is the density function of an {\em Exp}$(\theta)$ random variable, and that the prior $\pi(\theta)\propto\theta^{\alpha-1}e^{-\beta\theta}$ is given by a {\em Gamma}$(\alpha, \beta)$ distribution with shape and rate parameters $\alpha>0$ and $\beta>0$. Then from (\ref{easyex1}), and for $0<h<y_{obs}+\beta$, we can directly obtain
\begin{eqnarray*}
p_{ABC}(y_{obs}|\theta) & = & \frac{1}{2h}e^{-\theta y_{obs}}(e^{\theta h}-e^{-\theta h})\\
\hat{b}_h(y_{obs}|\theta) & = & \frac{1}{6}h^2\theta^3e^{-\theta y_{obs}}\\
\pi_{ABC}(\theta|y_{obs}) &=& \frac{
\theta^{\alpha-1}e^{-\theta(y_{obs}+\beta)}\left(e^{\theta h}-e^{-\theta h}\right)
}{
\frac{\Gamma(\alpha)}{(y_{obs}+\beta-h)^\alpha}-\frac{\Gamma(\alpha)}{(y_{obs}+\beta+h)^\alpha}
},
\end{eqnarray*}
where $\Gamma(\alpha)=\int_0^\infty z^{\alpha-1}e^{-z}dz$ is the gamma function.

\begin{figure}[tb]
\centering
\includegraphics[width=12cm]{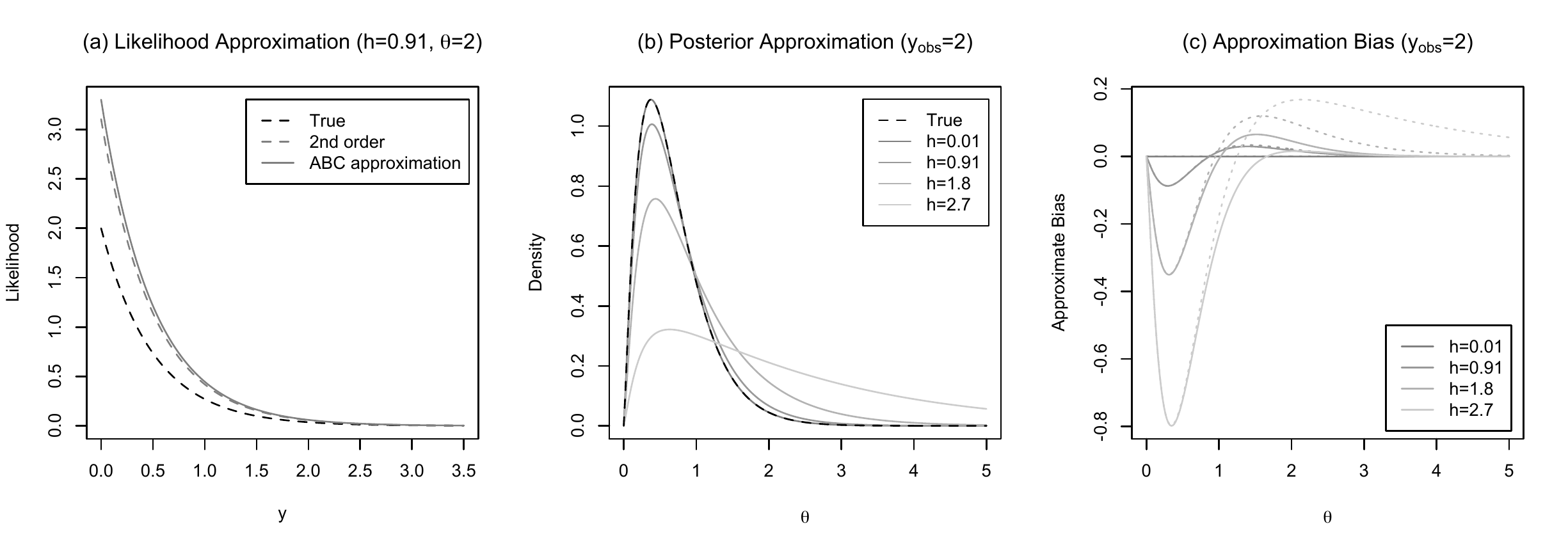}
\caption{\small
Approximations involved in the ABC analysis of the Exponential-Gamma example.
(a) Various likelihood functions with $h=0.91$ and $\theta=2$. The true likelihood function, $p(y|\theta)$, and the ABC approximation to the likelihood, $p_{ABC}(y|\theta)$, are denoted by black-dashed and solid grey lines respectively. The second order approximation to $p_{ABC}(y|\theta)$, given by $p(y|\theta)+\hat{b}_h(y|\theta)$, is illustrated by the grey-dashed line.
(b) The ABC posterior approximation, $\pi_{ABC}(\theta|y_{obs})$ with $y_{obs}=2$ for various values of $h=0.01, 0.91, 1.80, 2.70$.
(c) Approximation bias in the ABC posterior as a function of $h$ for $y_{obs}=2$. Dashed lines indicate the exact bias $a(\theta|y_{obs})$ for each $h$, whereas solid lines denote the second order bias $\hat{a}(\theta|y_{obs})$.
}
\label{chapter3:toy}
\end{figure}

Figure \ref{chapter3:toy}(a) illustrates the true likelihood function, $p(y|\theta)$, (black dashed line) and  the ABC approximation to the true likelihood function, $p_{ABC}(y|\theta)$, (solid grey line) as a function of $y$ for $h=0.91$ and $\theta=2$. Also shown (grey dashed line), is the second order approximation to the ABC likelihood function, $p(y|\theta)+\hat{b}_h(y|\theta)$. In this case, the second order approximation provides a reasonable representation of the ABC likelihood, $p_{ABC}(y|\theta)$. For other choices of $h$ and $\theta$, the quality of this representation will vary.

The ABC approximation, $\pi_{ABC}(\theta|y_{obs})$, to the true posterior $\pi(y_{obs}|\theta)$, given $y_{obs}=2$ and $\alpha=\beta=1.2$ is shown in Figure \ref{chapter3:toy}(b) for various values of $h=0.01, \ldots, 2.7$ (grey lines). The true posterior is illustrated by the black dashed line. For small $h$ ($h=0.01$), $\pi_{ABC}(\theta|y_{obs})$ is indistinguishable from the true posterior. As $h$ increases, so does the scale of the approximate posterior, which begins to exhibit a large loss of precision compared to the true posterior. Both mean and mode of $\pi_{ABC}(\theta|y_{obs})$ increase with $h$.

Finally, Figure \ref{chapter3:toy}(c) shows the resulting 
bias, $a_h(\theta|y_{obs})$, in the ABC posterior approximation as a function of $\theta$ and $h$. Dashed and solid lines respectively show the exact bias $a_h(\theta|y_{obs})$ and the second order bias $\hat{a}_h(\theta|y_{obs})$ (defined as $a_h(\theta|y_{obs})$ in (\ref{eqn:ahat}) but with $\hat{b}_h(y|\theta)$ substituted for $b_h(y|\theta)$).
 Clearly, the bias in the main body of the distribution, particularly in the region around the mode, is well described by the second order approximation, $\hat{a}_h(\theta|y_{obs})$, whereas the bias in the distributional tails is more heavily influenced by terms of higher order than two.
\\


%
%



\noindent{\bf Example 2:}\\
Suppose that the observed data, $y_{obs}=(y_{obs,1},\ldots,y_{obs,n})^\top$, are $n$ independent draws from a univariate ${N}(\theta,\sigma_0^2)$ distribution, where the standard deviation, $\sigma_0>0$, is known.
For this model we know that $p(y_{obs}|\theta)\propto p(\bar{y}_{obs}|\theta)$, where $\bar{y}_{obs}=\frac{1}{n}\sum_i y_{obs,i}$, as the sample mean is a sufficient statistic for $\theta$.
If we specify $K_h(u)$ as a Gaussian ${N}(0,h^2)$ kernel (see Table \ref{Chapter3:table:StandardKernels}), then the ABC approximation to the likelihood, $p(\bar{y}_{obs}|\theta)$ is given by
 \begin{eqnarray*}
	p_{ABC}(\bar{y}_{obs}|\theta)
	& = &  \int_{-\infty}^{\infty} K_h(|\bar{y}-\bar{y}_{obs}|)p(\bar{y}|\theta)d\bar{y}\nonumber\\
	& = &  \int_{-\infty}^\infty \frac{1}{\sqrt{2\pi}h}\exp\left\{-\frac{(\bar{y}-\bar{y}_{obs})^2}{2h^2}\right\}
	 \frac{\sqrt{n}}{\sqrt{2\pi}\sigma_0}\exp\left\{-\frac{n(\bar{y}-\theta)^2}{2\sigma_0^2}\right\} d\bar{y}\\
	& \propto & \exp\left\{  -\frac{(\theta-\bar{y}_{obs})^2}{2(\sigma^2_0/n+h^2)}\right\}
\end{eqnarray*}
for $h\geq0$.
That is, $\bar{y}_{obs}\sim{N}(\theta,\sigma_0^2/n+h^2)$ under the ABC approximation to the likelihood. In comparison to the true likelihood,  for which $\bar{y}_{obs}\sim{N}(\theta,\sigma_0^2/n)$,
the variance is inflated by $h^2$, the variance of the Gaussian kernel.
Accordingly, if the prior for $\theta$ is given by a ${N}(m_0,s_0^2)$ distribution, where $m_0$ and $s_0>0$ are known, then
\[
 \pi_{ABC}(\theta|y_{obs})=\phi\left(\frac{m_0s_0^{-2}+\bar{y}_{obs}(\sigma_0^2/n+h^2)^{-1}}{s_0^{-2}+(\sigma_0^2/n+h^2)^{-1}},
\frac{1}{s_0^{-2}+(\sigma_0^2/n+h^2)^{-1}} \right),
\]
where $\phi(a,b^2)$ denotes the density of a $N(a,b^2)$ distributed random variable.

Clearly $\pi_{ABC}(\theta|y_{obs})\rightarrow \pi(\theta|y_{obs})$ as $h\rightarrow 0$. However, the approximation will be quite reasonable if $\sigma^2/n$ is the dominating component of the variance so that $h$ is small in comparison \cite{drovandi12}.
A similar result to the above is available in the case of a multivariate parameter vector, $\theta$.

\begin{figure}[tb]
\centering
\includegraphics[width=12cm]{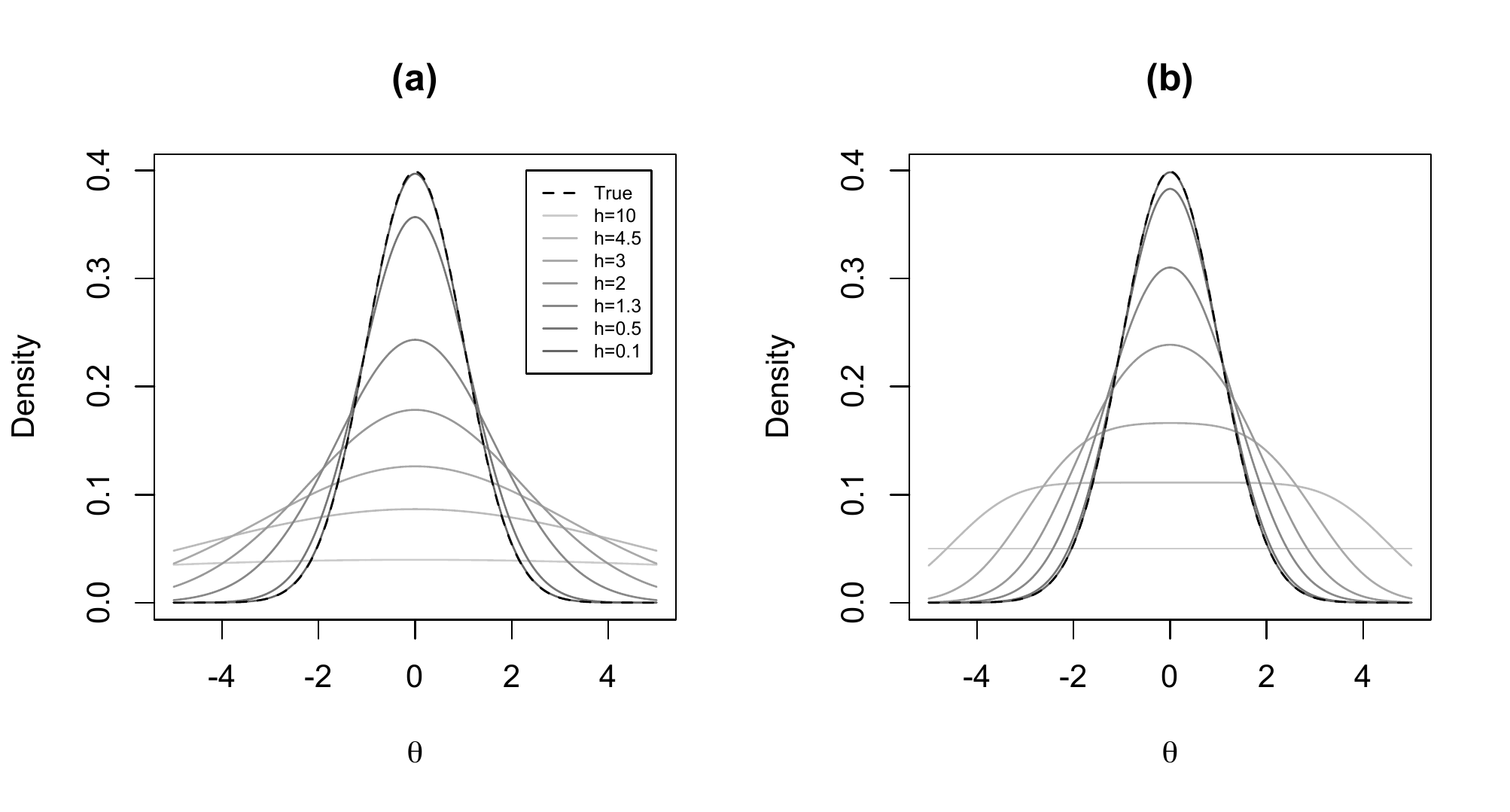}
\caption{\small
ABC posterior approximations, $\pi(\theta|y_{obs})$, for a $N(0,1)$ target distribution (dashed lines) for various values of kernel scale parameter $h$. The posterior approximations are based on (a) $N(0,h^2)$ and (b) uniform over $[-h,h]$ kernel functions, $K_h(u)$.
}
\label{chapter3:toy-normal}
\end{figure}

Figure \ref{chapter3:toy-normal}(a) illustrates the resulting ABC posterior approximation $\pi_{ABC}(\theta|y_{obs})$ with $\bar{y}_{obs}=0$ when $\sigma^2/n=1$ for the improper prior given by $m_0=0, s^2_0\rightarrow\infty$, so that the true posterior distribution is $N(0,1)$ (dashed line). The approximation is clearly quite reasonable for $h=0.5$ and $h=0.1$ as then $h^2<\sigma^2_0/n$.
Figure \ref{chapter3:toy-normal}(b) shows the same posterior approximations  but based on a uniform kernel over $[-h, h]$ for $K_h(u)$, rather than the Gaussian $N(0,h^2)$ kernel. This ABC posterior is derived from (\ref{easyex1}). The resulting forms for $\pi_{ABC}(\theta|y_{obs})$ are no longer within the Gaussian family for $h>0$, exhibit a flatter behaviour around the mean, and are more concentrated around the mean due to the compact support of the uniform kernel. The approximations with either kernel perform well for small $h$.

This example additionally provides some insight into the asymptotic behaviour of the ABC posterior approximation. Following standard likelihood asymptotic results, when the amount of data, $n$, becomes large, the true likelihood function, $p(y|\theta)$, will approximately behave as a Gaussian distribution. As most prior distributions will have little impact in this setting (they will be approximately constant over the region of high posterior density), it follows that the ABC posterior approximation, $\pi_{ABC}(\theta|y_{obs})$ will follow a Gaussian distribution with a variance that is inflated by an $h^2$ term. Consequently, the ABC posterior approximation, $\pi_{ABC}(\theta|y_{obs})$ may then in principle be improved simply by rescaling the posterior variance to remove this term \cite{drovandi12}.

\section{The use of summary statistics}

\subsection{Summary statistic basics}
\label{section:summaryStatisticBasics}

\noindent Despite the development in the previous Section, the ABC posterior approximation $\pi_{ABC}(\theta|y_{obs})\propto\int K_h(\|y-y_{obs}\|)p(y|\theta)p(\theta)dy$ is rarely used in practice. This is because, except in very specific scenarios (such as when $y_{obs}$ is very low dimensional, or when the likelihood function $p(y|\theta)$ factorises into very low dimensional components), it is highly unlikely that $y\approx y_{obs}$ can be generated from $p(y|\theta)$ for any choice of $\theta$ for realistic datasets. This results in the need to use a large value of the kernel scale parameter $h$ in order to achieve viable rejection sampling algorithm acceptance rates (or a similar loss of performance in other algorithms), and in doing so produce poorer ABC posterior approximations.

In the stereological extremes analysis in Section \ref{section:stereological} we replaced the full dataset $y_{obs}$ with a sufficient statistic $n_{obs}$ for the model parameter $\lambda$ when estimating $\pi(\theta|y_{obs})=\pi(\lambda|n_{obs})$. As sufficient statistics can be much lower dimensional than the full dataset, it is clear that greater approximation accuracy can be achieved for the same computational overheads when using low dimensional  statistics  (which is hinted at in the $g$-and-$k$ distribution analysis in Section \ref{section:gandk}).
The following example, based on \citeN{drovandi12}, highlights the computational benefits in using lower dimensional, and less variable sufficient statistics.
\\

\noindent {\bf Example 3:}\\
Suppose that $y=(y_1,y_2)^\top$,  where $y_i\sim${\em Binomial}$(n,\theta)$ with $\theta\sim U(0,1)$. Consider three possible vectors of sufficient statistics: $s^1=(y_1,y_2)^\top$ is the full dataset, $s^2=(y_{(1)},y_{(2)})^\top$ are the order statistics $y_{(1)}\leq y_{(2)}$, and $s^3=y_1+y_2$ is the sum of the two individual values. All three vectors of statistics are sufficient for this simple model.

It is easy to compute the marginal distribution of each summary statistic $p_i(s^i) = \int_0^1p(s^i|\theta)\pi(\theta)d\theta$ as follows:
\begin{eqnarray*}
	p_1(s^1) & = &\int_0^1\prod_{i=1}^2\left(\begin{array}{c}n\\y_i\end{array}\right)
				\theta^{y_i}(1-\theta)^{n-y_i} d\theta\\
			& = & \left(\begin{array}{c}n\\y_1\end{array}\right)\left(\begin{array}{c}n\\y_2\end{array}\right)
				B(y_1+y_2+1,2n-y_1-y_2+1),\\
	p_2(s^2) & = & \left[2-I(y_{(1)}=y_{(2)})\right]\int_0^1\prod_{i=1}^2\left(\begin{array}{c}n\\y_i\end{array}\right)
				\theta^{y_i}(1-\theta)^{n-y_i} d\theta\\
			& = & \left[2-I(y_{(1)}=y_{(2)})\right]
			 \left(\begin{array}{c}n\\y_1\end{array}\right)\left(\begin{array}{c}n\\y_2\end{array}\right)
				B(y_1+y_2+1,2n-y_1-y_2+1),\\
	p_3(s^3) & = & \int_0^1\left(\begin{array}{c}2n\\s^3\end{array}\right)
				\theta^{s^3}(1-\theta)^{2n-s^3} d\theta\\
			& = & \left(\begin{array}{c}2n\\s^3\end{array}\right)
				B(s^3+1,2n-s^3+1)\\
			& = & 1/(2n+1),
\end{eqnarray*}
where $\mbox{B}(a,b)=\int_0^1z^{a-1}(1-z)^{b-1}dz$ is the beta function. Here, $p_i(s^i)$ is the probability of generating the vector $s^i$ under an ABC rejection sampling algorithm with sampling distribution given by the prior, $g(\theta)=\pi(\theta)$. That is, $p_i(s_i)$ is the acceptance probability of the algorithm if we only accept those sufficient statistics that exactly match the observed sufficient statistics.

Suppose that we observe $y_{obs}=(y_{obs,1},y_{obs,2})^\top=(1,2)^\top$ from $n=5$ experiments. From the above, we have algorithm acceptance rates of:
\[
	p_1(s^1_{obs})=\frac{5}{132}\approx 0.038, \:\:\:\: p_{2}(s^2_{obs})=\frac{5}{66}\approx 0.076\quad\mbox{and}\:\:\:\: p_3(s^3_{obs})=\frac{1}{11}\approx 0.091,
\]
where $s^i_{obs}$ denotes the statistic $s^i$ derived from $y_{obs}$.
The probability $p_1(s^1)$  is the probability of generating first $y_1=1$  and then $y_2=2$. As a result, $p_1(s^1)$ will decrease rapidly as the length of the observed dataset $y_{obs}$ increases. The probability $p_2(s^2)$ corresponds to the probability of generating either $y=(1,2)^\top$ or $y=(2,1)^\top$, which are equivalent under the binomial model. Hence, $s^2$ has twice the probability of $s^1$ of occurring. Finally, the probability $p_3(s^3)$, is the probability of generating $y=(1,2)^\top, (2,1)^\top, (0,3)^\top$ or $(3,0)^\top$. Each of these cases are indistinguishable under the assumed model, and so the event $s^3$ occurs with the largest probability of all.

Quite clearly, while still producing samples from the true target distribution, $\pi(\theta|y_{obs})$, the impact on the efficiency of the sampler of the choice of sufficient statistics is considerable, even for an analysis with only two observations, $y_1$ and $y_2$.  The most efficient choice is the minimal sufficient statistic. The differences in the acceptance rates of the samplers would become even greater for larger numbers of observations, $n$.
\\

While the optimally informative choice of statistic for an ABC analysis is a minimal sufficient statistic, this may still be non-viable in practice. For example, if the minimal sufficient statistic is the full dataset $y_{obs}$, sampling from $\pi_{ABC}(\theta|y_{obs})$ will be highly inefficient even for moderately sized datasets. Similarly, in a scenario where the likelihood function may not be known beyond a data generation procedure, identification of any low-dimensional sufficient statistics (beyond, trivially, the full dataset $y_{obs}$) may be impossible. Further, low dimensional sufficient statistics may not even exist, depending on the model.

In general, a typical ABC analysis will involve specification of a vector of summary statistics $s=S(y)$, where $\dim(s)\ll\dim(y)$. The rejection sampling algorithm with then contrast $s$ with $s_{obs}=S(y_{obs})$, rather than $y$ with $y_{obs}$. As a result, this procedure will produce samples from the distribution $\pi_{ABC}(\theta|s_{obs})$ as follows:

\begin{table}
\caption{\bf ABC Rejection Sampling Algorithm}
\noindent {\it Inputs:}
\begin{itemize}
\item A target posterior density $\pi(\theta|y_{obs})\propto p(y_{obs}|\theta)\pi(\theta)$, consisting of a prior distribution $\pi(\theta)$ and a procedure for generating data under the model  $p(y_{obs}|\theta)$.
\item A proposal density $g(\theta)$, with $g(\theta)>0$ if $\pi(\theta|y_{obs})>0$.
\item An integer $N>0$.
\item A kernel function $K_h(u)$ and scale parameter $h>0$.
\item A low dimensional vector of summary statistics $s=S(y)$.
\\
\end{itemize}

\noindent {\it Sampling:}\\
\noindent For $i=1, \ldots, N$:
\begin{enumerate}
\item \label{chapter3:alg:ABC-rejectionSS:step1} Generate $\theta^{(i)}\sim g(\theta)$ from sampling density $g$.
\item Generate $y\sim p(y|\theta^{(i)})$ from the likelihood.
\item Compute summary statistic $s=S(y)$.
\item Accept $\theta^{(i)}$ with probability $\frac{K_h(\|s-s_{obs}\|)\pi(\theta^{(i)})}{Kg(\theta^{(i)})}$\\
	where $K\geq K_h(0)\max_\theta\frac{\pi(\theta)}{g(\theta)}$. Else go to \ref{chapter3:alg:ABC-rejectionSS:step1}.
\\
\end{enumerate}

\noindent {\it Output:}\\
A set of parameter vectors $\theta^{(1)},\ldots,\theta^{(N)}$ $\sim$ $\pi_{ABC}(\theta|s_{obs})$.
\end{table}

Similar to the discussion in Section \ref{chapter3:section:TheApproximatePosteriorDistribution}, it can be seen that the ABC posterior approximation now has the form
\begin{equation}
\label{ABCpostApproxSS}
	\pi_{ABC}(\theta|s_{obs}) \propto \int K_h(\|s-s_{obs}\|)p(s|\theta)\pi(\theta)ds,
\end{equation}
where $p(s|\theta)$ denotes the likelihood function of the summary statistic $s=S(y)$ implied by $p(y|\theta)$.
(That is, $p(s|\theta) = \int_\mathcal{Y}\delta_{s}(S(y))p(y|\theta) dy$.)
If we let $h\rightarrow 0$, so that only those samples, $\theta$, that generate data for which $s=s_{obs}$ are retained, then
\begin{eqnarray*}
	\lim_{h\rightarrow 0} \pi_{ABC}(\theta|s_{obs}) & \propto & \int \lim_{h\rightarrow 0}K_h(\|s-s_{obs}\|)p(s|\theta)\pi(\theta)ds\\
	& = & \int \delta_{s_{obs}(s)}p(s|\theta)\pi(\theta)ds\\
	& = & p(\theta|s_{obs})\pi(\theta).
\end{eqnarray*}
Hence, samples from the distribution $\pi(\theta|s_{obs})$ are obtained as $h\rightarrow 0$. If the vector of summary statistics, $s=S(y)$, is sufficient for the model parameters, then $\pi(\theta|s_{obs})\equiv\pi(\theta|y_{obs})$, and so samples are produced from the true posterior distribution.  However, if $S(y)$ is not sufficient -- and this is typically the case in practice -- then the ABC posterior approximation is given by (\ref{ABCpostApproxSS}), where in the best scenario (i.e. as $h\rightarrow 0$) the approximation is given by $\pi(\theta|s_{obs})$.

The following example illustrates the effect of using a non-sufficient summary statistic.
\\

\noindent{\bf Example 4:}\\
Consider again the univariate Gaussian model in Example 2. Suppose that we modify  this example \cite{drovandi12}, so that the model still assumes that the observed data $y_{obs}=(y_{obs,1},\ldots,y_{obs,n})^\top$ are random draws from a univariate $N(\theta,\sigma^2_0)$ distribution, but where we now specify an insufficient summary statistic, $s=\bar{y}_{1:n'}=\frac{1}{n'}\sum_{i=1}^{n'}y_i$ with $n'< n$.

Writing $s_{obs}=S(y_{obs})$, the resulting ABC approximation to the likelihood function becomes
 \begin{eqnarray*}
	p_{ABC}(s_{obs}|\theta)
	& = &  \int K_h(s-s_{obs})p(s|\theta)ds\\
	&\propto & \int_{-\infty}^\infty\frac{1}{\sqrt{2\pi}h}\exp\left\{-\frac{(s-s_{obs})^2}{2h^2}\right\} \frac{\sqrt{n'}}{\sqrt{2\pi}\sigma_0}\exp\left\{-\frac{n'(s-\theta)^2}{2\sigma^2_0}\right\}ds\\
	& \propto & \exp\left\{  -\frac{(\theta-s_{obs})^2}{2(\sigma^2_0/\omega n+h^2)}\right\},
\end{eqnarray*}
where $\omega=n'/n$ is the proportion of the $n$ observations used in the vector of summary statistics.
That is, $s_{obs}\sim N(\theta,\sigma^2/\omega n + h^2)$.  When $\omega=1$, then $s_{obs}=\bar{y}_{obs}$ is sufficient for $\theta$ and so $s_{obs}\sim N(\theta,\sigma^2/n + h^2)$ recovers the same result as Example 2.

When $n'<n$, so that $s$ is no longer sufficient for $\theta$, 
the mean of the Gaussian likelihood function is centred on the mean $\bar{y}_{obs,1:n'}$ rather than $\bar{y}_{obs,1:n}$, but more critically
the variance of the Gaussian likelihood is $\sigma^2/\omega n + h^2$. It is evident that there are now two sources of error, both of which inflate the variance of the likelihood. The first, $h^2$, arises through the matching of the simulated and observed data through the Gaussian kernel. The second source of error comes from the $0<\omega<1$ term, which can be interpreted as the degree of inefficiency of replacing $y$ by $s=S(y)$. That is, the use of non-sufficient statistics reduces the precision of the likelihood (and by turn, the posterior distribution) in this case.

From Example 2, it follows that when $n$ is large and the posterior is asymptotically Gaussian, the ABC posterior approximation, $\pi_{ABC}(\theta|s_{obs})$, can be improved by rescaling to remove $h^2$ from the posterior variance. However, correcting for the lack of sufficiency in the summary statistic, $s$, would require knowledge of the relative inefficiency of $s$ over $y$, which may be difficult to obtain in practice.
\\

The choice of summary statistics for an ABC analysis is a critical decision that directly affects the quality of the posterior approximation. Many approaches for determining these statistics are available, and these are reviewed in \shortciteN{blum+nps13} and \citeN{prangle17}, this volume. These methods seek to trade off two aspects of the ABC posterior approximation that directly result from the choice of summary statistics. The first is that $\pi(\theta|y_{obs})$ is approximated by $\pi(\theta|s_{obs})$. As this represents an irrevocable potential information loss, the information content in $s_{obs}$ should be high. The second aspect of the ABC posterior approximation is that the simulated and observed summary statistics are compared within a smoothing kernel $K_h(\|s-s_{obs}\|)$ as part of the form of $\pi_{ABC}(\theta|s_{obs})$ (\ref{ABCpostApproxSS}). As stochastically matching $s$ and $s_{obs}$ becomes increasingly difficult as the dimension of the summary statistics increases, the dimension of $s$ should be low.

As such, the dimension of the summary statistic should be large enough so that it contains as much information about the observed data as possible, but also low enough so that the curse-of-dimensionality of matching $s$ and $s_{obs}$ is avoided. For illustration, in Example 3, the optimum choice of summary statistic is a minimal sufficient statistic. However, for other models it may be the case that the dimension of the minimal sufficient statistic is equal to that of the original dataset. As this will cause curse-of-dimensionality problems in matching $s$ with $s_{obs}$, it is likely that a more accurate ABC posterior approximation can be achieved by using a lower-dimensional non-sufficient statistic, rather than remaining within the class of sufficient statistics. This was indeed the case in the $g$-and-$k$ distribution analysis in Section \ref{section:gandk}.

\subsection{Some practical issues with summary statistics}

Even with the above principles in mind, summary statistic choice remains one of the most challenging aspects of implementing ABC in practice. For instance, it is not always viable to continue to add summary statistics to $s$ until the resulting ABC posterior approximation does not change for the worse, as is illustrated by the following example.
\\

\noindent {\bf Example 5:}\\
Suppose that  $y=(y_1,\ldots,y_n)^\top$ with $y_i\sim${\em Poisson}$(\lambda)$. Combined with conjugate prior beliefs $\lambda\sim${\em Gamma}$(\alpha,\beta)$ this gives $\lambda|y\sim${\em Gamma}$(\alpha+n\bar{y},\beta+n)$. For this model we know that the sample mean $\bar{y}$ is a sufficient statistic. However, we also know that the mean and variance of a {\em Poisson}$(\lambda)$ model are both equal to $\lambda$, and so we might also expect the sample variance $v^2$ to also be informative for $\lambda$, although it is not sufficient. Suppose that we observe $y_{obs}=(0,0,0,0,5)^\top$ which gives $(\bar{y}_{obs},v^2_{obs})=(1,5)$. Here, as the sample mean and variance are quite different from each other, we might expect that the Poisson model is not appropriate for these data.

\begin{figure}[tb]
\centering
\includegraphics[width=12cm]{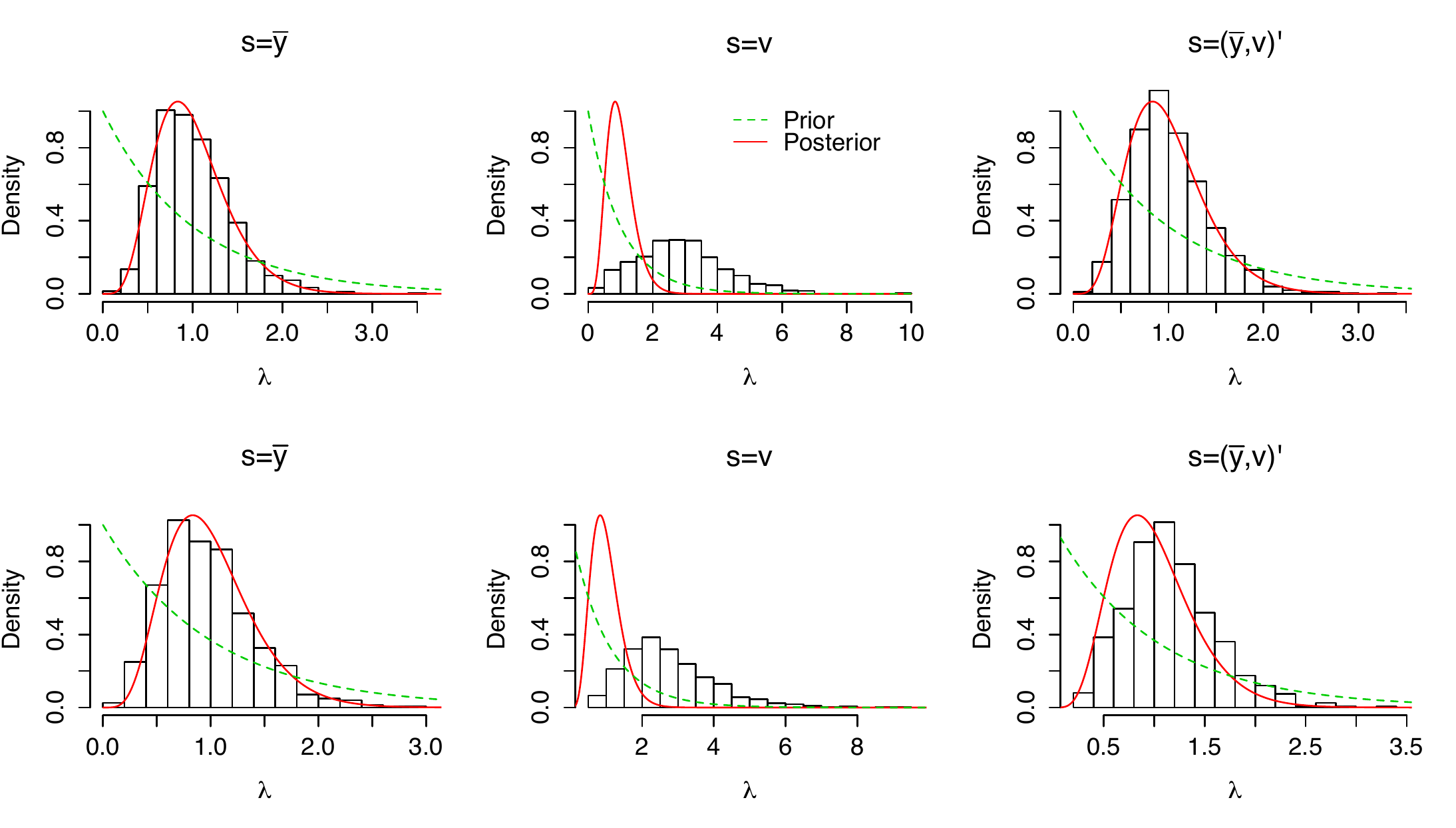}
\caption{\small
Various ABC posterior approximations (histograms) for a {\em Gamma}$(\alpha+\bar{y},\beta+n)$ target distribution (solid line) with a {\em Gamma}$(\alpha,\beta)$ prior  (dashed lines).
Columns illustrate posterior estimates based on (left) sample mean $s=\bar{y}$,  (centre) standard deviation $s=v$  and (right) $s=(\bar{y},v)^\top$ as summary statistics. Top row shows results with $h=0$ and the bottom row with $h=0.3$.}
\label{chapter3:bias}
\end{figure}

Figure \ref{chapter3:bias} illustrates various ABC posterior approximations to the true target distribution (solid lines)
based on a prior with $\alpha=\beta=1$ (dashed lines), with
$K_h(u)$ specified as a uniform kernel over $[-h,h]$ and $\|u\|$ representing Euclidean distance.
The top row illustrates the resulting posterior approximations, $\pi(\lambda|s_{obs})$, when the summary statistics $s$ are given as the sample mean $\bar{y}$ (left panel), the sample standard deviation $v$ (centre), or both (right) when the kernel scale parameter  is $h=0$. Using $s=\bar{y}$ recovers the true posterior exactly, which is no surprise as $\bar{y}$ is a sufficient statistic. Using $s=v$ produces an informed ABC approximation, but one which is based on a variance that is consistent with a larger mean under the Poisson model. When $s=(\bar{y},v)^\top$ then we again obtain the true posterior distribution as $\pi(\lambda|\bar{y}_{obs},v_{obs})\equiv \pi(\lambda|\bar{y}_{obs})$ through sufficiency, and the additional information that $v$ brings about the sample $y$ has no effect on the ABC estimated posterior.

The bottom row in Figure \ref{chapter3:bias} shows the same information as the top row, except that the kernel scale parameter is now non-zero ($h=0.3$). The posterior approximations based on $s=\bar{y}$ and $s=v$ are minor deviations away from those in the top row when $h=0$. This occurs as the values of $\lambda$ that are able to reproduce the observed summary statistics within a non-zero tolerance $h=0.3$ are slightly different to those that can reproduce the summary statistics exactly. However, the third panel with $s=(\bar{y},v)^\top$ is clearly biased to the right, with the resulting ABC posterior approximation visually appearing to be a loose average of those
distributions with $s=\bar{y}$ and $s=v$.

This behaviour is different from when $h=0$. In that case, when adding more information in the vector of summary statistics in going from $s=\bar{y}$ to $s=(\bar{y},v)^\top$, the posterior approximation does not change as the summary statistic $s=\bar{y}$ is sufficient and it is being matched exactly. However, when $h>0$, because the ABC algorithm allows a non perfect matching of the sufficient statistic $\bar{y}$, it additionally allows the extra information in the sample standard deviation $v$ to also contribute to the approximation. In this case, because the observed summary statistics $\bar{y}_{obs}$ and $v_{obs}$ are inconsistent with respect to the model, this then results in a strongly biased fit when moving from $s=\bar{y}$ to $s=(\bar{y},v)^\top$. 

As such, while it may be tempting to include progressively more summary statistics into $s_{obs}$ until the ABC posterior approximation does not change appreciably, the assumption that that this will provide the most accurate posterior approximation is clearly incorrect. Even if $s_{obs}$ contains sufficient statistics for the model, the inclusion of further statistics can still bias the posterior approximation, particularly in the case where the observed data are inconsistent with the model.
\\

The identification of suitable summary statistics is clearly a critical part of any analysis. Accordingly many techniques have been developed for this purpose -- see e.g. \shortciteN{blum+nps13} and \citeN{prangle17} (this volume) for a detailed review and comparison of these methods. While the choice of summary statistics is itself of primary importance, it is less appreciated that the distance measure $\|\cdot\|$ can also have a substantial impact on ABC algorithm efficiency, and therefore the quality of the posterior approximation.

\begin{figure}[tb]
\centering
\includegraphics[width=7cm]{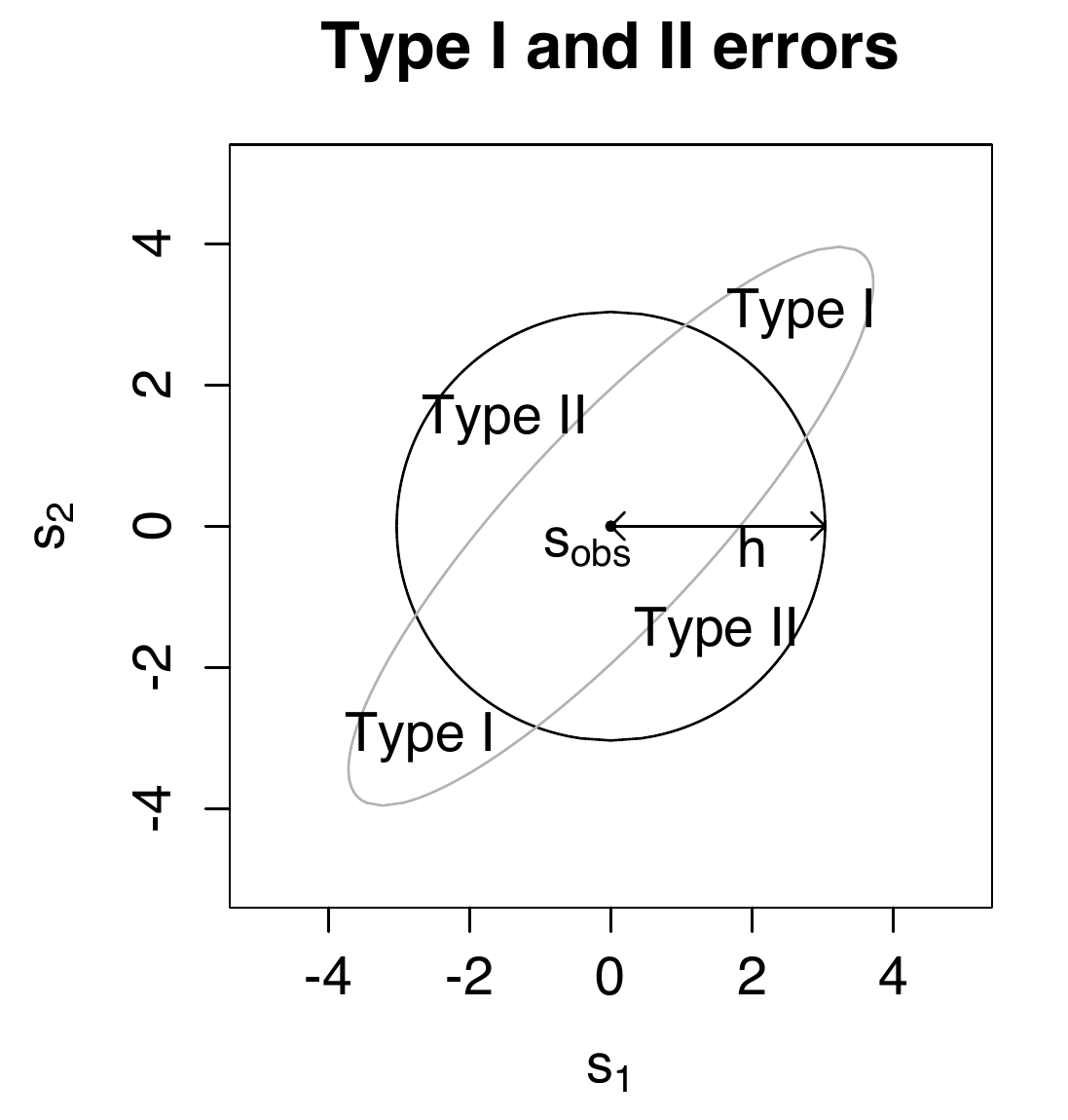}
\caption{{\protect\small The concept of type I and II errors for accept/reject decisions in ABC samplers under a uniform kernel, $K_h(u)$, over $[-h,h]$ and Euclidean distance, $\|\cdot\|$. The circle represents the acceptance region for a simulated summary statistic $s=(s_1,s_2)^\top$, centred on $s_{obs}$. The ellipse represents the possible dependence between $s_1$ and $s_2$.
}}
\label{image:type12}
\end{figure}

Consider the distance measure  $\|s-s_{obs}\|=(s-s_{obs})^\top\Sigma^{-1}(s-s_{obs})$. Here we can specify the covariance matrix $\Sigma$ as the identity matrix to produce Euclidean distance, or as a diagonal matrix of non-zero weights to give weighted Euclidean distance (e.g. \shortciteNP{hamilton+crhbe05,luciani+sjft09}) or as a full covariance matrix to produce Mahalanobis distance (e.g \shortciteNP{peters+fs12,erhardt+s16}).
To see why standard and weighted Euclidean distance can be a poor choice, consider the setting in Figure \ref{image:type12}, where candidate parameter values, $\theta$, generating continuous bivariate statistics, $s|\theta$, $s=(s_1,s_2)^\top$, are accepted as draws from $\pi_{ABC}(\theta|s_{obs})$ if $s$ lies within a ball of radius $h$, centered on $s_{obs}$.   That is, $K_h$ is the uniform kernel on $[-h,h]$, and $\|\cdot\|$ denotes Euclidean distance. 

If we reasonably suppose that the elements of $s$ may be dependent and on different scales, their true distribution under the model may be better represented by an ellipse (grey lines). As such, an efficient ABC algorithm should accept candidate draws from $\pi_{ABC}(\theta|s_{obs})$ if $s|\theta$ lies within this ellipse.
Consequently, implementing a circular acceptance region (implying independence and identical scales) induces both type I (i.e. candidate samples are rejected when they should be accepted) and type II (i.e.  candidate samples are accepted when they should be rejected) errors.

Work linking the ABC posterior with non-parametric density estimation methods (\citeNP{blum10}; see Section \ref{section:interpretations}) provides support for this argument.
Here, for a multivariate kernel $K_H(u)=\det(H)^{-1}K(H^{-1}u)$, where $K$ is a symmetric multivariate density function with zero mean and finite variance, a general rule of thumb is to specify the bandwidth matrix as
$H\propto\Sigma^{-1/2}$ where $\Sigma$ is the covariance matrix of the data (e.g. \citeNP{scott92,wand+j95}). In the ABC context, this is equivalent  to defining $\|\cdot\|$ as Mahalanobis distance where $\Sigma$ is the covariance matrix of $s$ (or $s|\theta$).

Note that the above argument assumes that the summaries $s_1$ and $s_2$ are both informative for the model parameter $\theta$. For example, in the case where $s_1+s_2$ is uninformative, but $s_1-s_2$ is informative, then it is credible that the circular acceptance region could result in a more accurate ABC posterior approximation than that resulting from the elliptical region. In general, the best acceptance region is tied up with the choice of the summary statistics in a more complicated way than that presented here (see e.g. \citeNP{prangle17b} for a discussion).

The following example illustrates the effect that different covariance matrices $\Sigma$ can have on the ABC posterior approximation.
\\

\noindent {\bf Example 6:}\\
Suppose that the model is specified as $y_1,\ldots,y_{50}\sim N(\theta,1)$, with a uniform prior $\theta\sim U(-5,5)$. Various sufficient statistics are available for this model. We consider two alternatives: $s^1=(\bar{y}_{1:40},\bar{y}_{41:50})^\top$ and $s^2=(\bar{y}_{1:25}-\bar{y}_{26:50}, \bar{y}_{26:50})^\top$ where $\bar{y}_{a:b}=(b-a+1)^{-1}\sum_{i=a}^by_i$. In each case, given the observed sufficient statistics $s_{obs}=(0,0)^\top$, the exact posterior distribution $\pi(\theta|y_{obs})$ is $N(0,1/50)$ truncated to $(-5,5)$. However, the covariance matrices of $s^1$ and $s^2$ for fixed $\theta$ are quite different
(though they do not depend on the exact value of $\theta$),
namely
\begin{equation}
\label{eqn:13}
	 \mbox{Cov}(s^1|\theta)=\left(\begin{array}{cc}1/40&0\\0&1/10\end{array}\right),
	\quad
	 \mbox{Cov}(s^2|\theta)=\left(\begin{array}{cc}\phantom{-}2/25&-1/25\\-1/25&\phantom{-}1/25\end{array}\right),
\end{equation}
with a negative correlation between the elements of $s^2$ of $-1/\sqrt{2}\approx-0.71$.
We implement ABC using the distance measure as $\|s-s_{obs}\|=(s-s_{obs})\Sigma^{-1}(s-s_{obs})'$ and consider the impact of the choice of $\Sigma$.

We use a version of the ABC rejection sampling algorithm (see box) that maintains a sample $\theta^{(1)}, \ldots, \theta^{(N)}$ of size $N$ from the ABC posterior approximation, which progressively lowers the kernel scale parameter $h$ until a stopping rule is satisfied. On algorithm termination, the samples are identical to those samples that would have been obtained under the standard ABC rejection sampling algorithm if it was implemented with the lowest value of $h$ achieved under the stopping rule. This allows us to implement a rejection sampling algorithm that will terminate when a pre-specified degree of accuracy has been achieved. The (random) number of iterations obtained before algorithm termination will accordingly be an indicator of the efficiency of the model specification -- in this case, the effect of different covariance matrices $\Sigma$.

\begin{table}
\caption{\bf ABC Rejection Sampling Algorithm (with Stopping Rule)}
\noindent {\it Initialise:}\\
For each particle $i=1,\ldots,N$:
\begin{itemize}
	\item Generate $\theta^{(i)}\sim\pi(\theta)$ from the prior, $y^{(i)}\sim p(y|\theta^{(i)})$ from the likelihood.
	\item Compute summary statistics $s^{(i)}=S(y^{(i)})$, and distance $\rho^{(i)}=\|s^{(i)}-s_{obs}\|$.
	\item Generate $u^{(i)}\sim\mbox{U}(0,1)$ that determines whether to accept the particle. \\(i.e. accept if $u^{(i)}\leq K_h(\rho^{(i)})/K_h(0)$.)
	\item Determine the {\it smallest} $h$ that results in the acceptance of all $N$ particles. E.g.
	\[
		 h=\sqrt{\max_i\{-[\rho^{(i)}]^2/(2\log(u^{(i)}))\}}\qquad\mbox{or}\qquad
		h=\max_i\{\rho^{(i)}\}
	\]
	if (respectively)
	\[
		 K_h(\rho)\propto\exp\{-\rho^2/(2h^2)\}\qquad\mbox{or}\qquad
		K_h(\rho)\propto1\:\:\mbox{on } [-h,h].
	\]
	\item Calculate the acceptance probabilities $W^{(i)}=K_h(\rho^{(i)})/K_h(0)$, $i=1,\ldots,N$.
\end{itemize}

\noindent {\it Simulation:}\\
\noindent While the stopping rule is not satisfied, repeat:
\begin{enumerate}
	\item Identify the index of the particle that will first be rejected if $h$ is reduced: $r=\arg_i\min\{W^{(i)}-u^{(i)}\}$.
	\item Set the new value of $h$ to be the lowest value which would result in the acceptance of all particles, except particle $r$.
	\item Recompute acceptance probabilities $W^{(i)}$  given the new value of $h$.
	\item Replace particle $r$ by repeating:
	\begin{enumerate}
		\item Generate $\theta^{(r)}\sim\pi(\theta)$, $y^{(r)}\sim p(y|\theta^{(i)})$, $u^{(r)}\sim U(0,1)$.
		\item Compute $s^{(r)}=S(y^{(r)})$, $\rho^{(r)}=\|s^{(r)}-s_{obs}\|$,\\ $W^{(r)}=K_h(\rho^{(r)})/K_h(0)$
	\end{enumerate}
	Until $u^{(r)}\leq W^{(r)}$.
\end{enumerate}

\noindent {\it Output:}\\
A set of parameter vectors $\theta^{(1)},\ldots,\theta^{(N)}$ $\sim$ $\pi_{ABC}(\theta|s_{obs})$, with $h$ determined as the largest achieved value that satisfies the stopping rule.
\end{table}

Table \ref{table:stoppingrule} displays the average number of data generation steps (i.e. generating $y\sim p(y|\theta)$) in each algorithm implementation, per final accepted particle, as a function of smoothing kernel type and the form of $\Sigma$, based on 100 replicate simulations of $N=500$ samples. The stopping rule continued algorithm execution until an estimate of the absolute difference between empirical ($F_N(\theta)$) and true ($F(\theta)$) model cumulative distribution functions was below a given level. Specifically
when $\sum_{i=1}^N|F_N(\theta^{(i)})-F(\theta^{(i)})|<0.01825$.
In Table \ref{table:stoppingrule}, the true form of $\Sigma$ is given by $\mbox{Cov}(s^1|\theta)$ and $\mbox{Cov}(s^2|\theta)$ (\ref{eqn:13}), and the diagonal form refers to the matrix constructed from the diagonal elements of $\mbox{Cov}(s^2|\theta)$.

\begin{table}[tb]
\centering
\begin{tabular}{cl|cc|cc|cc}
Summary &&\multicolumn{6}{c}{Form of $\Sigma$}\\
Statistic & Kernel & \multicolumn{2}{c}{Identity} & \multicolumn{2}{|c|}{Diagonal} & \multicolumn{2}{c}{True}\\
\hline
& Uniform & 134.7  &  \phantom{1}(5.8) &&& \phantom{1}84.5 & (2.4) \\
$s=s^1$ & Epanechnikov & 171.6 &  \phantom{1}(4.7) &&& 111.1 & (3.8) \\
& Triangle & 232.3 &  \phantom{1}(7.1) &&& 153.0 & (5.1) \\
& Gaussian &  242.4&  \phantom{1}(6.5) &&& 153.6 & (4.9) \\
\hline
& Uniform & 182.5 &  \phantom{1}(5.6) & 161.0 & (4.1) & \phantom{1}84.4 & (2.4) \\
$s=s^2$ & Epanechnikov & 245.5 &  \phantom{1}(6.6) & 209.2 & (7.2) & 111.1 & (3.8)\\
& Triangle & 336.3 &  \phantom{1}(8.9) & 277.2 & (6.9) & 144.2 & (3.8)\\
& Gaussian & 368.2 & (12.6)  & 289.7  & (9.7)  &157.7  & (4.3)  \\
\end{tabular}
\caption{{\small Mean number of summary statistic generations per final accepted particle (with standard errors in parentheses), as a function of the form of covariance matrix, $\Sigma$, and smoothing kernel $K_h$, and for two different sets of sufficient statistics $s^1=(\bar{y}_{1:40},\bar{y}_{41:50})^\top$ and $s^2=(\bar{y}_{1:25}-\bar{y}_{26:50}, \bar{x}_{26:50})^\top$.  Results are based on 100 replicates of posterior samples of size $N=500$.
}}
\label{table:stoppingrule}
\end{table}

The summary statistics for $s=s^1$ are independent, but are on different scales. Accordingly, when this difference of scale is accounted for ($\Sigma =$ true), algorithm efficiency, and therefore ABC posterior approximation accuracy, is greatly improved compared to when the difference in scale is ignored ($\Sigma =$ identity). The summary statistics $s^2$ are both negatively correlated and on different scales. As for $s^1$, when summary statistic scale is taken into consideration ($\Sigma =$ diagonal) an improvement in algorithm efficiency and ABC posterior approximation accuracy is achieved compared to when it is ignored. However in this case, further improvements are made when the correlation between the summary statistics is also accounted for ($\Sigma=$ true). These results are consistent regardless of the form of the smoothing kernel $K_h$. Note that the uniform kernel produces the most efficient algorithm and most accurate ABC posterior approximation, and that this steadily worsens as the form of the kernel deviates away from the uniform density, with the worst performance is obtained under the Gaussian kernel.

This approach has been implemented in practice by e.g. \shortciteN{luciani+sjft09} and \citeN{erhardt+s16}, who identify some value of $\theta=\theta^*$ in a high posterior density region via a pilot analysis, and then estimate $\mbox{Cov}(s|\theta^*)$ based on repeated draws from $p(s|\theta^*)$.

\section{An ABC analysis in population genetics}

To illustrate some of the points concerning summary statistics we consider here a population genetic example, very similar to that considered in the paper by \shortciteN{pritch99}, a key paper in the development of ABC methods. In population genetics we are often confronted with sequence data (as illustrated in Table \ref{table:pg-data1}), and we wish to infer demographic parameters that may be associated with such data. The standard modelling framework that is used is Kingman's coalescent \shortcite{hein04}, which describes the genealogical relationship of DNA sequences in a sample. The general likelihood problem that we wish to solve then can be represented as
$$
p(y_{obs} | \phi) = \int_H p(y_{obs} | H)p(H | \phi) dH
$$
where $y_{obs}$ represents the observed set of sequences in a sample, $\phi$ is an unobserved vector of parameters, and $H$ represents the unobserved genealogy history, including mutations. A common mutation model, used here, is the infinite-sites model, in which every mutation that occurs in a genealogy is unique. Typically $H$ is high dimensional, represented as a variable-length vector of times of events in the genealogical history, and the types of events. Although the likelihood can be computed exactly for simple demographic models and small data sets \shortcite{hein04} it is generally more flexible to resort to Monte Carlo methods \cite{marj06}.

One approach is through importance sampling. Here, an instrumental distribution $q_{\phi,y}(H)$ is available that describes the distribution of all genealogical histories $H$ that are consistent with the data $y$, as a function of the model parameters $\phi$. The distribution $q_{\phi,y}(H)$ is easy to simulate from and has a known functional form that can be directly evaluated. It also has the property that $p(y|H')=1$ for $H'\sim q_{\phi,y}(H)$.
Hence, $p(y_{obs}|\phi)$ can be estimated by
$$
\hat{p}(y_{obs} | \phi) = \frac{1}{N}\sum_{i=0}^{N}\frac{p(H^{(i)}|\phi)}{q_{\phi,y_{obs}}(H^{(i)})}
$$
where $H^{(i)}\sim q_{\phi,y_{obs}}(H)$ for $i=1,\ldots,N$.

In this analysis we compare an ABC approach to the above importance sampling method that targets the true likelihood. The aim is to investigate the performance of different summary statistics on ABC inferences, using the importance sampling-based inferences as a (noisy) ground-truth. The demographic model that generates the data is one of smooth exponential expansion. In this model the current population size $N_0$ contracts backwards in time as $N_0(t) = N_0\exp(-\beta t)$ where time $t$ is expressed in units of $2N_0$ and $\beta = 2N_0b$ is the growth rate in this scaled time. An additional parameter in the model is the scaled mutation rate $\theta_0 = 4N_0\mu$. 

\begin{center}
\begin{table}[tb]
\centering
{\small \tt
1 : 000000000000000000000001000100000000000000\\
1 : 000000000000000000001010001000000000101001\\
1 : 000000000000000100000010001000010000101001\\
5 : 000000100000100000000000000000000000000000\\
1 : 000000100000100000000000000000001000000000\\
2 : 000000100000100000000000000001000000000000\\
1 : 000000100000100000000000000010000000000000\\
2 : 000000100000100001000000000000000000000000\\
2 : 000000100000100010000000000000000000000000\\
1 : 000000100001100001000000000000000000000000\\
1 : 000000100100100000100000000000000001000000\\
1 : 000000100100100000110000000000000000000000\\
1 : 000000101000100000000100100000000000000110\\
2 : 000001100010010000000000000000000000000000\\
2 : 000010010000000000000001010000000000010000\\
2 : 000100000000000000000000001000100000001000\\
1 : 001000000000001000000000001000000110101000\\
1 : 010000100000100000000000000000000000000000\\
2 : 100000000000000000000010001000000000101001\\
}
\caption{{\small Infinite sites data simulated with \emph{ms} in a format suitable for the \emph{Genetree} program. The left hand column gives the number of times the sequence on the right is observed in the sample (of size 30 in this case). The ancestral type is denoted by 0 and the mutant (derived) type is denoted by 1. The length of the sequence is equal to the number of segregating sites $S$ and is equal to the number of mutations that occurred in the genealogy. All sequences that share a mutation at a given position are descendent (and possibly further mutated) copies of the sequence in which that mutation first occurred. The sequences are ordered lexicographically.}}
\label{table:pg-data1}
\end{table}
\end{center}

In the ABC analysis, simulations are carried out using the \emph{ms} program of  \citeN{hudson02}. A technical complication that needs to be accounted for when using \emph{ms} is that time in this program is scaled in units of $4N_0$ rather than $2N_0$ that appears standardly in most treatments (\emph{e.g.} \shortciteNP{hein04}), and, more importantly, in the \emph{Genetree} importance sampling program  \cite{griff94} that is used for the ground-truth. The data in Table \ref{table:pg-data1} were generated using the \emph{ms} command:
\begin{verbatim}
ms 20 1 -t 50 -G 30
\end{verbatim}
which simulates one instance of 20 sequences with $\theta=50$ and $\alpha = 30$, where $\alpha = \beta/2$ (because of the different scaling of time, noted above). Assuming independent uniform priors $U(0,200)$ for each parameter $\phi=(\theta_0,\alpha)^\top$, it is straightforward to generate particles by sampling parameter values from the prior and then compute an importance weight for each particle using an algorithm suggested by \citeN{stephens00}. The implementation here (described in \citeNP{maciuca12}) is a modification of the \emph{Genetree} program 
to include the Stephens and Donnelly algorithm, following \citeN{deiorio04}. Although the particles could be used directly for weighted density estimation, it is computationally easier to first resample them in proportion to their weights $w^{(i)}$, because the distribution of weights is typically very skewed (they have high variability). For the data in Table \ref{table:pg-data1}, $N=10^8$ generated particles yielded an effective sample size
 (estimated by $(\sum_i w^{(i)})^2/\sum_i w^{(i)2}$) of around $300$. The following analyses are based on resampling 1000 particles.

For the ABC analysis, parameter values $\phi=(\theta_0,\alpha)^\top$ are simulated from the prior, data sets are simulated using {\em ms}, and summary statistics computed. The four summary statistics examined comprise the number of segregating sites, $S_0$, which corresponds to the number of mutations in the genealogy under the infinite sites mutation model, the average pairwise Hamming distance between all pairs of sequences in the sample, $\pi_0$, Tajima's $D$, 
and Fay and Wu's $H_0$. 
These latter two statistics express the difference in estimates of the scaled mutation parameter $\theta_0$, assuming a standard coalescent model (\emph{i.e.} with no population growth), based on two different unbiased estimators, one of which is $\pi_0$. The average pairwise distance, $\pi_0$, is directly an estimate of $\theta_0$ because in the standard constant size model the expected time to coalescence for a pair of sequences is $2N_0$, and therefore the expected number of mutations occurring down both branches since the common ancestor is $(2N_0 + 2N_0)\mu$. Other estimators have been developed, based on the number of segregating sites (Watterson's estimator, used in Tajima's $D$), or the number of segregating sites weighted by the number of times the mutant type occurs in the sample (Fu's estimator, used in Fay and Wu's $H_0$). Only under the standard constant size model will these estimators all have the same expectation, and therefore deviations between them can be used to identify departures from this model. Negative values of $D$ and positive values of $H_0$ are expected to be found in growing populations. The output of the \emph{ms} program can be piped to a program \verb"sample_stats", included with \emph{ms}, which computes these four summary statistics.
The observed summary statistics are:
\[
s_{obs}=(\pi_0,S_0,D,H_0)^\top= (5.90, 42,-1.64,3.67)^\top.
\]

ABC methods were implemented by first simulating $N=1,000,000$ parameter values from the $U(0,200)$ prior distributions, storing these in the file \texttt{params.txt}
(in the order indicated by the key-word \texttt{tbs})  and then running the \emph{ms} program with the command
\begin{verbatim}
ms 20 1 -t tbs -G tbs < params.txt
\end{verbatim}
The summary statistics corresponding to these simulated data were then obtained and then $\|s-s_{obs}\|$ computed as Euclidean distance. The ABC posterior approximation was obtained by using a uniform kernel $K_h$ over $[-h,h]$ and determining the kernel scale parameter $h$ as the value retaining the 1000 samples for which $s^{(i)}$ is closest to $s_{obs}$.

\begin{figure}[tbh]
\centering
\includegraphics[width=10cm]{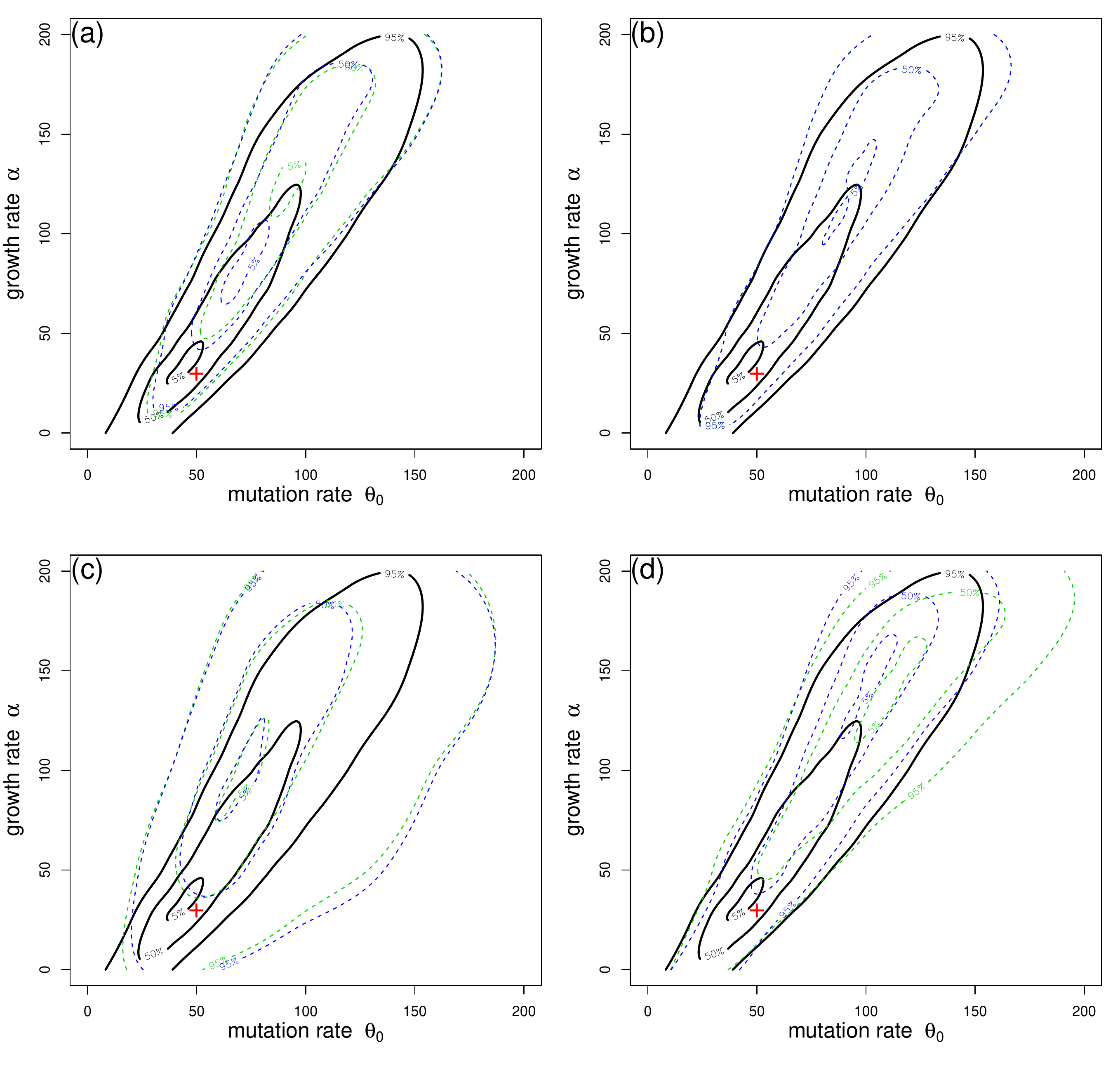}
\caption{{\protect\small Various ABC posterior approximations using different summary statistics and scalings, compared to the `ground-truth' importance sampling based posterior (black lines). 
The true parameter value is indicated by a $+$.
Estimates show the 95\%, 50\%, and 5\% highest posterior density contours. 
ABC posteriors are based on (a) all four summary statistics; (b) $\pi_0$ and $S_0$ only; (c) $D$ and $H_0$ only; (d) $\pi_0$ (green dotted) and $S_0$ (blue dotted). 
For panels (a)-(c) the ABC posterior is based on scaled summary statistics (blue dotted line), and unscaled summary statistics (green dotted line).
}}
\label{image:pg-simfig}
\end{figure}

The summary statistics are measured on different scales. A common practice is to centre and scale them using the standard deviation for each summary statistic sampled from the prior predictive distribution.
(However, some authors argue that the motivations for this are flawed as an arbitrary change in the prior can change the scaling of a summary statistic within the analysis. Instead, following a similar discussion to that in Example 6, the scaling should be based on $\mbox{Cov}(s|\theta^*)$ for some value of $\theta=\theta^*$ in the high posterior density region, rather than $\mbox{Cov}(s)$. See e.g. \citeNP{erhardt+s16}.)
For the present analysis, the prior predictive sample standard deviations  for $\pi_0$, $S_0$, $D$ and $H_0$ are 14.3, 69.0, 0.50 and 7.3 respectively.
In Figure \ref{image:pg-simfig}  the estimated posterior distributions using both scaled and unscaled summary statistics are shown.

Figure \ref{image:pg-simfig} compares the resulting ABC posterior approximation using (a) all four summary statistics, (b) $D$ and $H_0$ only, (c) $\pi_0$ and $S_0$ only, or (d) $\pi_0$ or $S_0$ alone.
The first point to note is that the data, although quite informative about $\theta_0$ or $\alpha$ jointly, do not allow us to make very detailed inference about either parameter individually \emph{i.e.} they are only partially identifiable in the model -- at least for these data. This is the case both for the full-likelihood and ABC inferences, although the density for the full-likelihood method, as estimated by importance sampling, tends to be more localised towards the true parameter value (indicated by a $+$). 

When all four summary statistics are used (panel a) the 95\% HPD envelope for ABC is quite similar to that for importance sampling (black line), but is shifted towards higher values of $\alpha$ and $\theta_0$. Scaled or unscaled summary statistics give similar results. The ABC posterior approximation for $\pi_0$ and $S_0$ together (panel b) is very similar to that for the full set of summary statistics. In this case the distances for scaled and unscaled summaries are the same because $S$ is discrete and matched exactly. This outcome perhaps indicates that one should be cautious of adding summaries such as Tajima's $D$ because it is simply a nonlinear function of $\pi_0$ and $S_0$. Whereas $H_0$ includes additional information from the site frequency spectrum, and would be expected to be informative (positive $H_0$ indicates a deficit of high-frequency derived mutations compared with that expected under the standard model). Using $D$ and $H_0$ together (panel c) yields a less concentrated posterior approximation. Both statistics are based on the difference of two estimators of mutation rate, and therefore it is unsurprising that $\theta_0$ is not well localised. The posteriors based on $\pi_0$ and $S_0$ individually (panel d) superficially look surprisingly similar to the full-likelihood posterior. However there is much stronger support for larger values of $\theta_0$ and $\alpha$ than in the importance-sampling based posterior.

\begin{table}[tb]
\centering
{\small \tt
1 : 000000000000000000000000000000000010100001\\
1 : 000000000000000000000000001000000000000010\\
1 : 000000000000000000000001010100111001000100\\
4 : 000000000000000011010000000100000000000000\\
1 : 000000000000000111010010000100000100000000\\
4 : 000000000000000111010010000101000100000000\\
1 : 000000000000010000000000000000000000000000\\
1 : 000000000000100111010000000100000000000000\\
1 : 000000000001000000000000010100000000001100\\
1 : 000000000010000000000000010100000000000100\\
1 : 000000000100001000000100100000000010000000\\
1 : 000000010000000000000000000000000010100001\\
1 : 000100001000000000000000000100000000000100\\
1 : 001000000001000000101000010100000000010100\\
1 : 010001100000000000000000001000000000000000\\
1 : 100010000000000011010000000110000000000000\\
}
\caption{{\small Data from locus 9pMB8 surveyed in 11 Biaka pygmies (Hammer et al. 2010), 
using the same layout as for Table \ref{table:pg-data1} }}
\label{table:pg-data2}
\end{table}
We conduct a similar analysis  with sequence data published in \shortciteN{hammer10}  from locus 9pMB8 surveyed in 11 Biaka pygmies (resulting in 22 sequences). The data are shown in Table \ref{table:pg-data2}.  Like the simulated data above, there are 42 sites that are segregating within the Biaka sample and which are compatible with the infinite sites model. The ABC simulations were performed as previously, using all four summary statistics. The observed summary statistics for these data are
\[
s_{obs}=(\pi_0,S_0,D,H_0)^\top= (7.52,42,-1.35,4.0)^\top.
\]
The posterior computed using importance sampling was also computed as before, but required $12 \times 10^8$ particles to achieve a similar effective sample size to that for the previous data set. 

\begin{figure}[tb]
\centering
\includegraphics[width=5cm]{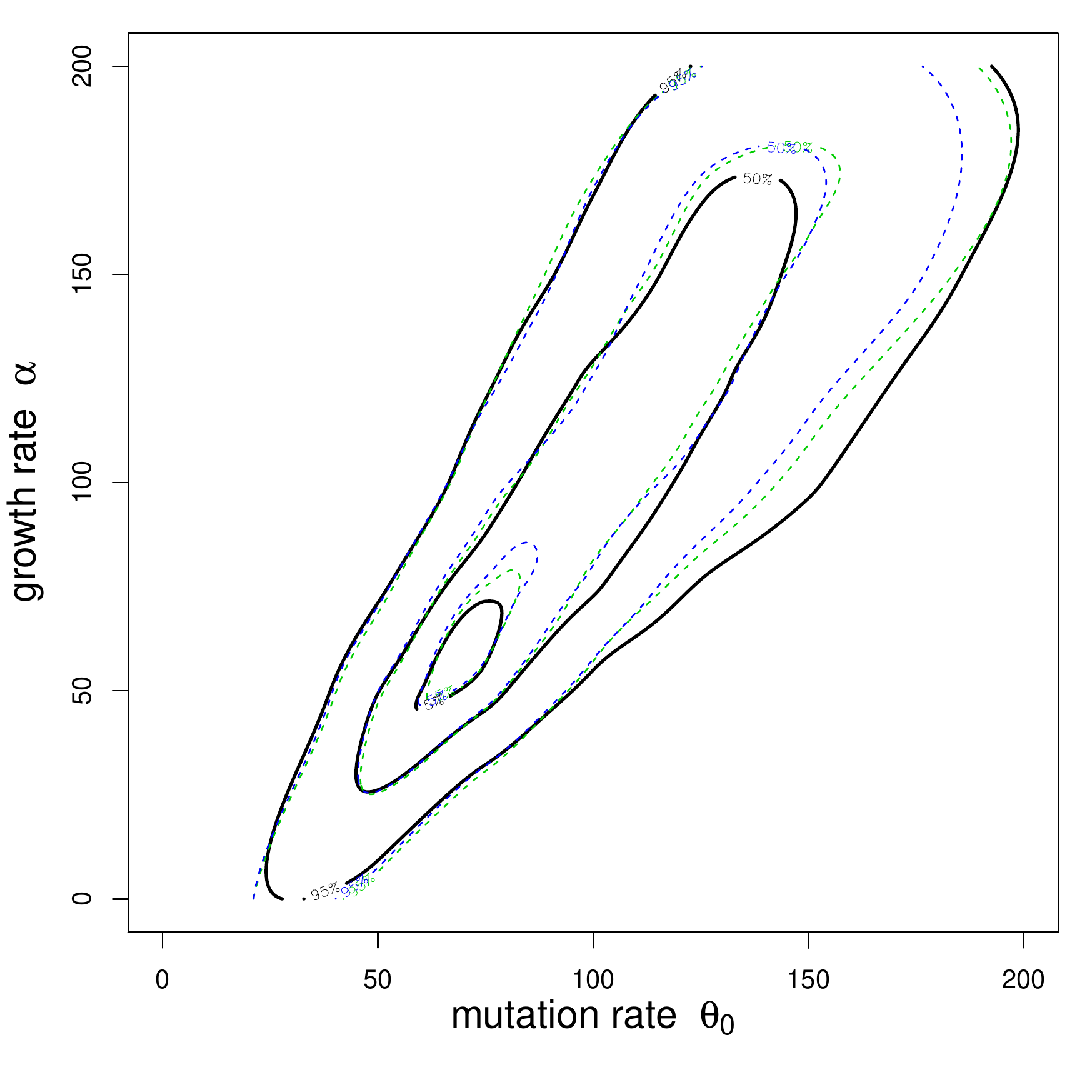}
\caption{{\protect\small Comparison of ABC posterior approximations (dotted lines) and full-likelihood (black lines) posterior for the Biaka pygmy data in Table \ref{table:pg-data2}. ABC posterior approximations are based on all four summary statistics which, are scaled (blue dotted line) and unscaled (green dotted line). 
}}
\label{image:pg-datafig}
\end{figure}

It is immediately apparent from Figure \ref{image:pg-datafig} that the ABC posterior approximation and ground-truth posterior are very similar, unlike the previous analysis. This differing behaviour is not due to Monte Carlo error.
The result illustrates a point that outside the exponential family there is no single, low-dimensional set of summary statistics $s$ that will be highly informative for $\theta$, for all observed datasets. Summary statistics that work well for one dataset may perform less well on another. In the case of the two datasets considered here, it may be argued that in the latter, despite the smaller sample size, there is a stronger signal of growth in these data, which is more readily captured by the summary statistics. For the simulated data the signal is less strong, and information in other summary statistics, such as the site frequency spectrum, or higher moments of the distribution of pairwise Hamming distances, may be required for the ABC posterior to better match the true posterior.

From a computational perspective, the $10^6$ ABC simulations took about 3 minutes on a desktop computer, whereas $10^8$ importance sampling simulations took around 4 hours i.e. the computational effort per iteration is broadly similar for both approaches. The algorithms used in each are `similar yet different', in that they both generate genealogical trees, but in one case the tree is constrained by the data, and in the other it is independent of the data. Naively one might think that an importance sampling  algorithm should be more efficient because it always generates a tree that is compatible with the data. However, it is typically very difficult to devise an algorithm that samples trees in proportion to their conditional distribution under the model, and therefore genealogical importance sampling tends to be inefficient, as illustrated here, where $10^8$ simulations only give an effective sample size of around 300. Of course, it is possible to use sequential methods, or a pseudo-marginal method to improve efficiency (\shortciteNP{andrieu+lv17,cornuet12,beaumont03}), but similar approaches are available for ABC as well.

\section{Levels of approximation in ABC}

The primary challenge in implementing an ABC analysis is to reduce the impact of the approximation, while restricting the required computation to acceptable levels. In effect this is the usual ``more computation for more accuracy'' tradeoff. It is therefore worthwhile to briefly summarise  the quality and nature of the approximations involved in any ABC analysis.
While some of these approximations are common with standard Bayesian analyses, in particular points \ref{chapter3:approxList1} and  \ref{chapter3:approxList5} below, within the ABC framework these have additional, more subtle implications.
In order, from model conception to implementation of the analysis, the ABC approximations are:
\begin{enumerate}
\item \label{chapter3:approxList1} {\em All models are approximations to the real data-generation process.}

While this is true for any statistical analysis, this approximation can produce an ABC-specific issue if the assumed model is not sufficiently flexible to be able to reproduce the observed summary statistics.
In this scenario
the kernel scale parameter $h$ will necessarily be large (as all simulated data are far from the observed data), and as a consequence the quality of the ABC approximation may be low.  Further, if, for this inflexible model, the observed summary statistics contain conflicting information for a model parameter, this may cause additional bias in the posterior approximation for this parameter, as is illustrated in Example 5. In summary, this means that the more unlikely a model is to have generated the observed data, the worse the ABC approximation will be.
In general this is problematic, as it implies that routine inspection of the fitted ABC posterior may not in itself be enough to determine model adequacy, as the ABC posterior may be a poor estimate of the true posterior, and poor data generation models may appear more likely (with $h>0$) than they actually are (with $h=0$).
By extension, this also implies that posterior model probabilities of inadequate models (constructed from the normalising constant of the poorly estimated ABC posterior distribution)  may also be affected, although this has yet to be fully explored in the literature. See \citeN{fearnhead18}, for an exploration of related ABC asymptotics results to date, and \shortciteN{marin+per18} for particular methods for performing ABC model choice.

\item \label{chapter3:approxList2} {\em Use of summary statistics rather than full datasets.}

The full posterior distribution $\pi(\theta|y_{obs})\propto p(y_{obs}|\theta)\pi(\theta)$ is replaced by the partial posterior $\pi(\theta|s_{obs})\propto p(s_{obs}|\theta)\pi(\theta)$ where $s_{obs}=S(y_{obs})$ is a vector of summary statistics. If $S$ is sufficient for $\theta$, then there is no approximation at this stage. More commonly, for non-sufficient $S$, there is a loss of information.

\item  \label{chapter3:approxList3} {\em Weighting of summary statistics within a region of the observed summary statistics.}

The partial posterior $\pi(\theta|s_{obs})$ is replaced by the ABC approximation to the partial posterior
\[
	\pi_{ABC}(\theta|s_{obs})\propto \pi(\theta)\int K_h(\|s-s_{obs}\|)p(s|\theta)ds
\]
where $K_h$ is a standard smoothing kernel with scale parameter $h\geq 0$. If $h=0$ or in the limit as $h\rightarrow 0$ then there is no further approximation at this stage. In most cases however, $h>0$ and so ABC makes use of a kernel density estimate as an approximation to the true likelihood function. This aspect of approximation can be a particular problem in ABC when the number of model parameters $\theta$ is large, as then the vector of summary statistics, $s$, must be equivalently large for parameter identifiability, and hence the comparison $\|s-s_{obs}\|$ will suffer from the curse of dimensionality.

\item  \label{chapter3:approxList4} {\em Approximations due to other ABC techniques.}

There are a number of other ABC techniques not discussed in this Chapter that are optionally implemented in ABC analyses in order to improve some aspect of the approximations in points 1 and 2, or to achieve a greater computational performance. Many of these are discussed in later Chapters, but some common methods involve  post-processing techniques such as regression and marginal adjustments (e.g. \shortciteNP{beaumont+zb02,blum+f10,blum+nps13,blum17,nott+ofs17}),
or develop alternative approximations to the intractable likelihood function, while remaining in the ABC framework, such as Expectation-Propagation ABC, synthetic likelihoods, and copula or regression-density estimation models (e.g. \shortciteNP{barthelme+c14,barthelme+cc17,wood10,price+dln17,drovandi+mr17,li+nfs15,fan+ns13,nott+ofs17}).

\item  \label{chapter3:approxList5} {\em Monte Carlo error.}

In common with most Bayesian analyses, performing integrations using Monte Carlo methods introduces Monte Carlo error. Typically this error may be reduced by using larger numbers of samples from the posterior, or by reducing the variability of importance weights. The same is true for an ABC analysis, although with the additional point that more posterior samples effectively allows for a lower kernel scale parameter $h$ and consequently an improved ABC posterior approximation.
As a result,
for a fixed number of Monte Carlo samples, the choice of kernel scale parameter represents a typical bias-variance tradeoff: if $h$ is large, more posterior draws are available, reducing variance, but at the cost of a poorer ABC approximation; if $h$ is small, the ABC posterior approximation is improved, but Monte Carlo variance is increased.

\end{enumerate}

\section{Interpretations of ABC}
\label{section:interpretations}

There are a number of closely related ways in which ABC methods may be understood or interpreted.
The most common of these is conditional density estimation of the posterior (e.g. \shortciteNP{blum10,bonassi+yw11,nott+ofs17}) in the sense usually understood in a conventional Bayesian analysis. Before observing the data, the  distribution $\pi(\theta,y)=p(y|\theta)\pi(\theta)$ describes prior beliefs about the model parameters and credible datasets under the model. When a dataset $y_{obs}$ is observed, interest is then in the conditional distribution of $\theta$ given that $y=y_{obs}$. In the ABC setting, $\pi(\theta,y)$ is represented by the joint sample $(\theta^{(i)},y^{(i)})\sim\pi(\theta,y)$, $i=1,\ldots, N$. Weighting the vectors $\theta^{(i)}$ based on the value of $\|y^{(i)}-y_{obs}\|$ (larger weights for smaller $\|y^{(i)}-y_{obs}\|$), then produces an empirical  conditional density estimate of $\pi(\theta|y_{obs})$.

Similarly, we have already discussed that the ABC approximation to the true likelihood, $p_{ABC}(y_{obs}|\theta)$, is a kernel density estimate of $p(y|\theta)$, following (\ref{Chapter3:eqn:ABClikelihood}) and (\ref{eqn:conDenEst}). This allows ABC to be considered as a regular Bayesian analysis with an approximated likelihood function.

\citeN{fearnhead+p12} noted that the ABC approximation to the posterior can be considered as a continuous mixture of posterior distributions
\begin{eqnarray*}
	\pi_{ABC}(\theta|y_{obs}) & \propto & \int K_h(\|y-y_{obs}\|)p(y|\theta)\pi(\theta) dy\\
	& = & \int w(y)\pi(\theta|y) dy
\end{eqnarray*}
where $\pi(\theta|y)=p(y|\theta)\pi(\theta)/\pi(y)$, with weight function $w(y)\propto K_h(\|y-y_{obs}\|)\pi(y)$. This is the continuous equivalent of equation (\ref{eqn:discreteMixturePost}) obtained during the analysis of stereological extremes in Section \ref{sec:extremesAnalysis}.

While ABC is most often thought of as an approximate method, \citeN{wilkinson13} pointed out that ABC methods can be considered as exact if $e=y-y_{obs}$ (or $e=\|y-y_{obs}\|$) is considered as the error (either from observation error or model misspecification) obtained in fitting the model $p(y|\theta)$ to the observed data $y_{obs}$. From this perspective, the smoothing kernel $K_h$ is simply the density function of this error, so that $e\sim K_h$, and $h$ is a scale parameter to be estimated.

Finally, while ABC methods are universally used for the analysis of models with computationally intractable likelihood functions, it is often overlooked that they also provide a useful inferential mechanism for tractable models. 
As an illustration, consider a scenario where a standard Bayesian analysis is available for a complex, but incorrect model, given the observed dataset. Under this model, predictions of some particular quantity of interest, $T(y)$, could be precise, but completely implausible due to the limitations in the model. Consider now an ABC analysis based on this model, based on matching summary statistics that include $T(y)$. ABC methods would identify those parameter values $\theta$ that are most likely to have produced these statistics under the model. This means that predictions of $T(y)$ under the ABC approximation now have some chance of being accurate (although they may be less precise), as the model may be able to predict the summary statistics, including $T(y)$, even if it can't accurately predict the full dataset.
This allows ABC to be interpreted as a mechanism for fitting models based on summary statistics that may in fact be more useful than the exact inference with the full dataset. An explicit example of this in the robust model selection  context was given by \shortciteN{li+nfs15}.

Related arguments allow ABC to be thought of as a natural method to fit models when the full dataset ($y_{obs}$) is only partially observed ($s_{obs}$) and has missing data (see e.g. \shortciteNP{rodrigues+fst18}). ABC methods have also been used to determine weakly informative prior distributions in a regular tractable Bayesian analysis, exploiting the mechanism of predictive data matching to identify a priori non-viable regions of the parameter space \shortcite{nott+dme17}.

\section{Further reading}
\label{section:FurtherReading}

ABC methods have been extensively and rapidly developed since their first modern appearance in \shortciteN{tavare+bgd97} and \shortciteN{pritch99}. Naturally a number of review articles have been written for various discipline audiences to review the techniques available at the time. While with time such reviews can rapidly become dated, they often provide useful perspectives on ABC methods as viewed at the time.  See, for example, the reviews by
\citeN{beaumont10},
\shortciteN{bertorelle+bm10},
\shortciteN{blum+nps13},
\shortciteN{csillery+bgf10},
\citeN{sisson+f11},
\shortciteN{marin+prr12},
\citeN{turner+v12},
\citeN{robert16},
\citeN{erhardt+s16},
\shortciteN{lintusaari+gdkc16}
and
\citeN{drovandi17}.
Each of the chapters in this Handbook also makes for excellent reading and review material on focused aspects of ABC \shortcite{tavare17,blum17,fan+s18,prangle17,marin+per18,drovandi18,nott+ofs17,andrieu+lv17,fearnhead18,ratmann+chc18,drovandi+gkr18,wegmann18,barthelme+cc17}.

Because ABC methods are now recognised as a standard Bayesian tool, their scientific reach has effectively become as extensive as standard Bayesian methods. While it is accordingly futile to exhaustively describe all areas in which ABC has applied, the below selection is provided to provide a flavour of the impact ABC methods have had. Beyond the applications in this Handbook, ABC methods have been successfully applied to applications in
$\alpha$-stable models \shortcite{peters+sf12},
archaeology \shortcite{wilkinson+t09},
cell biology \shortcite{johnston+smbr14,vo+dpp15,vo+dps15},
coalescent models \shortcite{fan+k11,tavare+bgd97},
ecology \cite{jabot+c09,wood10},
evolutionary history of mosquitos \shortcite{bennett+slmkdahihplw16},
filtering \shortcite{jasra+smm12},
extreme value theory \shortcite{erhardt+s12,erhardt+s16},
financial modelling \shortcite{peters+sf12},
host-parasite systems \shortcite{baudet+dscgms15},
HIV contact tracing \shortcite{blum+t10},
human evolution \shortcite{fagundes+rbnsbe07},
hydrological models \shortcite{nott+fms14},
infectious disease dynamics \shortcite{luciani+sjft09,aandahl+rst12},
infinite mixture models for biological signalling pathways \shortciteN{koutroumpas+bvc16},
image analysis \shortcite{nott+fms14},
long range dependence in stationary processes \shortcite{andrade+r15},
operational risk \shortcite{peters+s06},
quantile distributions \shortcite{allingham+km09,drovandi+p11},
pathogen transmission \shortcite{tanaka+fls06},
phylogeography \shortcite{beaumont+al10},
protein networks \shortcite{ratmann+ahwr09,ratmann+jhsrw07},
population genetics \shortcite{beaumont+zb02},
psychology \cite{turner+v12},
single cell gene expression \shortcite{lenive+ks16},
spatial point processes \cite{shirota+g16},
species migration \shortcite{hamilton+crhbe05},
state space models \shortcite{vakilzadeh+hba17},
stochastic claims reserving \shortcite{peters+fs12},
susceptible-infected-removed (SIR) models \shortcite{toni+wsis09},
trait evolution \shortcite{slater+hwjra12}
and
wireless communications engineering \shortcite{peters+nsfy10}.
Within this Handbook novel analyses can be found in  \shortciteN{peters18}, \shortciteN{rodrigues+fst18}, \shortciteN{stumpf18}, \shortciteN{estoup18}, \shortciteN{holden18}, \shortciteN{wood18} and \shortciteN{fan18}.

\section{Conclusions}

ABC methods are based on an inherently simple mechanism -- simulating data under the model of interest and comparing the output to the observed dataset. While more sophisticated ABC algorithms and techniques have subsequently been developed (and many of these are discussed in more detail in this Handbook), this core mechanic remains a constant.
It is this methodological simplicity that has made ABC methods highly accessible to researchers in across  many disciplines. We anticipate that this will continue in the future.

\section*{Acknowledgments} 

SAS is supported by the Australian Research Council under the Discovery Project scheme (DP160102544), and the Australian Centre of Excellence in Mathematical and Statistical Frontiers (CE140100049).

\bibliographystyle{chicago}
\bibliography{intro-chapter}
\thispagestyle{empty}
\end{document}